\documentclass[5p,twocolumn]{elsarticle} %
\usepackage{verbatim}
\usepackage{color}
\journal{Carbon}

\usepackage[]{subfigure}
\usepackage{placeins} 
\usepackage[english]{babel}  
\usepackage[utf8x]{inputenc}
\usepackage{graphicx}
\usepackage{booktabs}
\usepackage{tabularx}
\usepackage{caption}
\usepackage{amsmath}
\usepackage{amssymb}
\usepackage{multirow}

\begin{document}

\begin{frontmatter}

\title{Mechanical and thermal properties of Graphene Random nanofoams via Molecular Dynamics simulations}

\author[DICAM,TIPFA]{Andrea Pedrielli}

\author[TIPFA,PRAGA]{Simone Taioli}

\author[TIPFA]{Giovanni Garberoglio\corref{mycorrespondingauthor2}}
\cortext[mycorrespondingauthor2]{Corresponding author}
\ead{garberoglio@ectstar.eu}

\author[DICAM,LONDON,KET,BIO]{Nicola Maria Pugno\corref{mycorrespondingauthor1}}
\cortext[mycorrespondingauthor1]{Corresponding author}
\ead{nicola.pugno@unitn.it}

\address[DICAM]{Laboratory of Bio-Inspired and Graphene Nanomechanics - Department of Civil, Environmental and Mechanical Engineering, University of Trento, Italy}
\address[TIPFA]{European Centre for Theoretical Studies in Nuclear Physics and Related Areas (ECT*-FBK) and Trento Institute for Fundamental Physics and Applications (TIFPA-INFN), Trento, Italy}
\address[PRAGA]{Faculty of Mathematics and Physics, Charles University, Prague, Czech Republic}
\address[LONDON]{School of Engineering and Materials Science, Materials Research Institute, Queen Mary University of London, UK}
\address[KET]{Ket Lab, Edoardi Amaldi Foundation, Italian Space Agency, Italy}

\begin{abstract}

Graphene foams have recently attracted a great deal of interest for their possible use in technological applications, such as electrochemical storage devices, wearable electronics, and chemical sensing. In this work, we present computational investigations, performed by using molecular dynamics with reactive potentials, of the mechanical and thermal properties of graphene random nanofoams.
In particular, we assess the mechanical and thermal performances of four families of random foams characterized by increasing mass density and decreasing average pore size. We find that the foams' mechanical performances under tension cannot be rationalized in terms of mass density, while they are principally related to their topology. Under compression, higher-density foams show the typical slope change in the stress--strain curve at $5-10$~\% strain, moving from linear elasticity to bending stress plateau. At variance, lower density foams display a quasi-linear behaviour up to $35$~\% strain.
Furthermore, we assess the thermal conductivity of these random foams using the Green--Kubo approach. While foam thermal conductivity is affected by both connectivity and defects, nevertheless we obtain similar values for all the investigated families, which means that topology is the critical factor affecting thermal transport in these structures.

\end{abstract}

\begin{keyword}
Graphene foams\sep  Molecular Dynamics \sep Porosity \sep Stress-strain curve \sep Thermal conductivity
%\MSC[2008] 00-01\sep  99-00
\end{keyword}

\end{frontmatter}

\section{Introduction}
%%%%%%%%%%%%%%%%%%%%%%%%%%%%%%%%%
% What is present in literature?%
%%%%%%%%%%%%%%%%%%%%%%%%%%%%%%%%%

Recently, an increasing interest has been paid to nanoporous materials. Porosity, indeed, can strongly increase the surface-to-volume ratio and enhance the specific mechanical properties, such as the specific modulus and strength, with respect to bulk material. For example, a high surface-to-volume ratio is desirable for gas adsorption and separation \citep{Garberoglio2015}, while improving specific mechanical properties using carbon-based porous materials are of interest for building lightweight structural components \citep{taioli1}.

Moreover, after the discovery of novel bi-dimensional materials \citep{Novoselov2005}, such as the hexagonal allotrope of boron nitride (h-BN) and graphene, several investigations have been focused onto the search of unconventional 3D structures that inherit the outstanding electrical \citep{Gruneis,Gruneis2}, thermal and mechanical properties of their 2D counterpart in order to achieve specific requirements.

In particular, graphene shows excellent tensile properties, such as fracture strength ($\sigma \simeq$ 130 GPa) and Young’s modulus (E $\simeq$ 1 TPa) coupled with relatively low density due to its bidimensionality, and thus it is the best candidate material to be used in the synthesis of foam assemblies with superior properties.  
Graphene-based nanofoams can be synthesised by using CVD on metallic scaffold as well as nanoparticles assemblies \citep{Drieschner2016, Christian2017,Taioli2014} or  
chemically derived by reducing graphene oxide \citep{Tao2016}. In mechanical and thermal applications, critical parameters are the concentration of defects, the topology as well as the inter-flake contacts. Despite this technological interest, however only a few computational investigations have been performed to characterize their electronic, thermal and mechanical properties \citep{Alonso2012, Wu2013, Pedrielli2017}. 
In particular, mechanical properties of porous materials at microscale can be studied by the Ashby-Gibson theory, in which a unit  cell  approach  is
combined with dimensional analysis \citep{Ashby2006}. While this approach can be useful to perform dimensional analysis and deliver scaling laws of mechanical properties with respect to density, however the effective properties of porous materials are not often a simple function of porosity. At odds, they usually strongly depend on features at the nanoscale, where local atomic interactions start to play a crucial role, or on the presence of struts and of random pores with very special shapes that destroy structural periodicity. Furthermore, deformation mechanisms at the mesoscale can be very different at the nanoscale, where the fine details of graphene topology come into play, and a multiscale approach should be devised \citep{taioli1}. 
Additionally, it turns out that carbon-based nanoporous materials with random porosity distribution exhibit poor scaling of the mechanical properties with decreasing density, even more pronounced than that of metal and polymeric foams \citep{Qin2017}. 
However, nanoporous graphene foams easily outperform polymeric foams at high density and can compete with their high-performance rivals, such as the metal foams. Thus, the interest in studying these random porous structures for energy storage and damping devices remains high. 

Moreover, the high porosity of random foams suggests a concurrent application of these materials as thermal insulators. In particular, our goal is to assess the dependence of the thermophysical properties on pore density and size, and to compare thermal insulation performances of graphene-based 3D structures with other widely used carbon-based foams, such as polyurethane and metal foams.

\begin{figure*}[hbt]
\centering
\subfigure[]{
\includegraphics[width=0.30\textwidth]{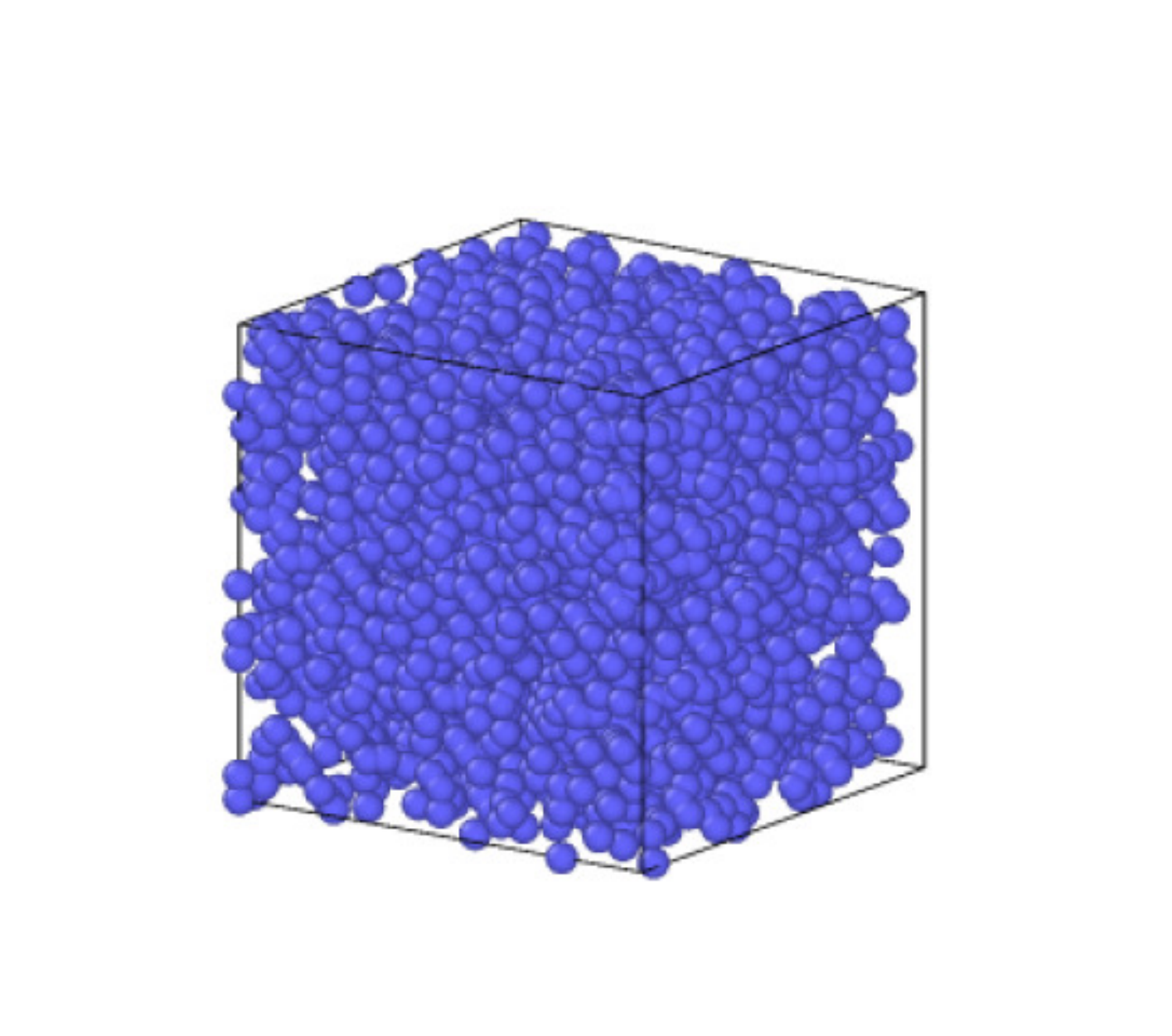}}
\subfigure[]{
\includegraphics[width=0.30\textwidth]{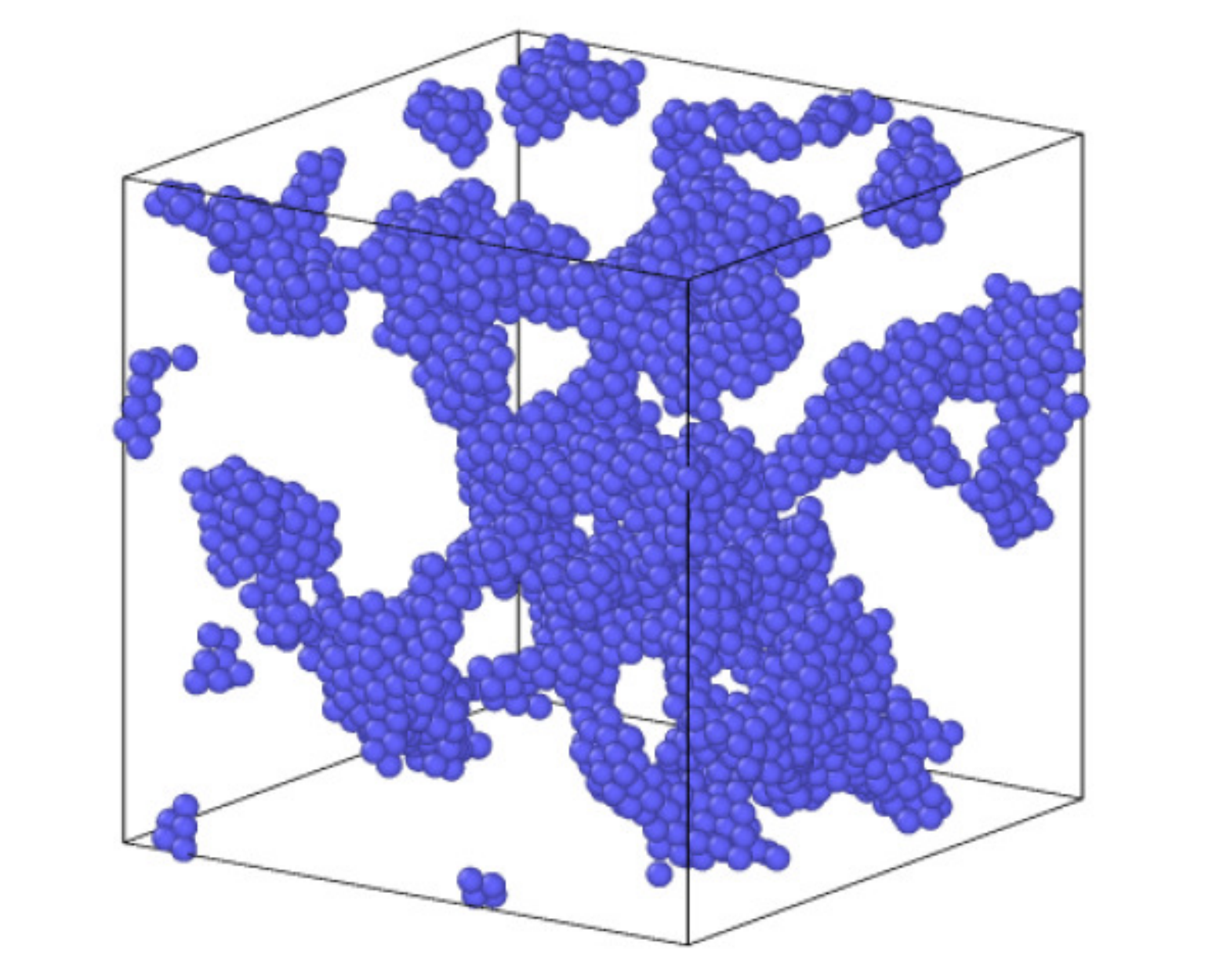}}
\subfigure[]{
\includegraphics[width=0.30\textwidth]{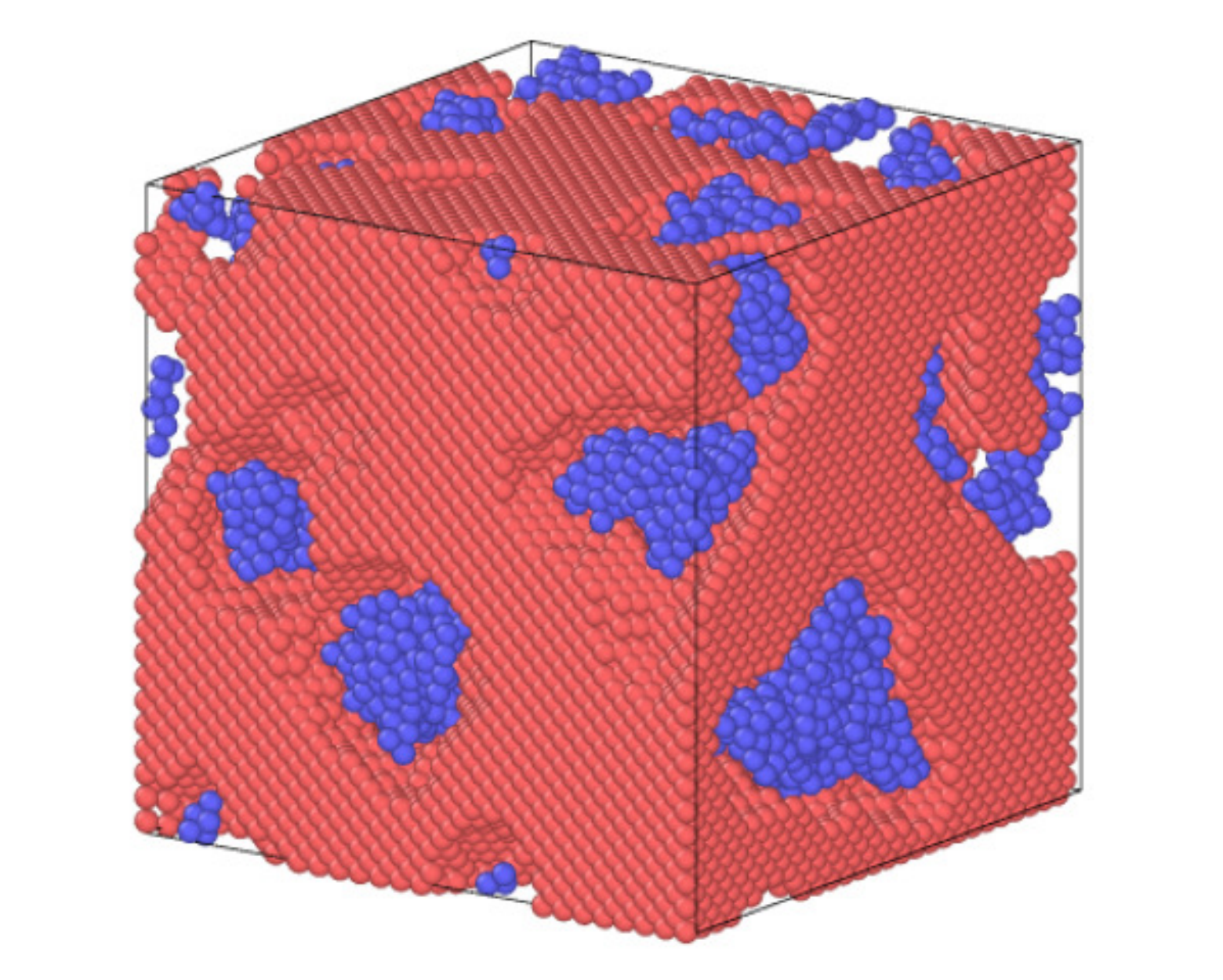}}
\subfigure[]{
\includegraphics[width=0.30\textwidth]{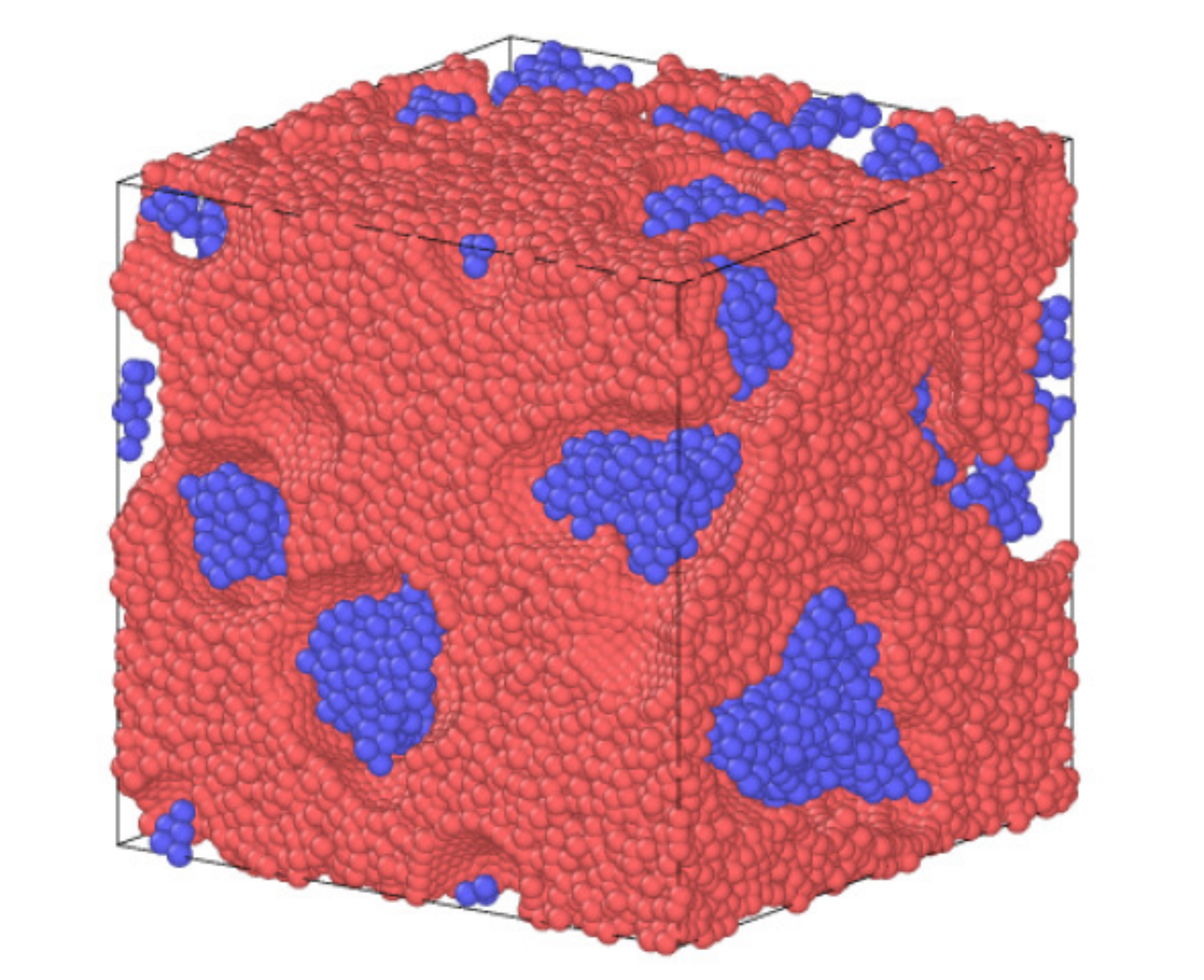}}
\subfigure[]{
\includegraphics[width=0.30\textwidth]{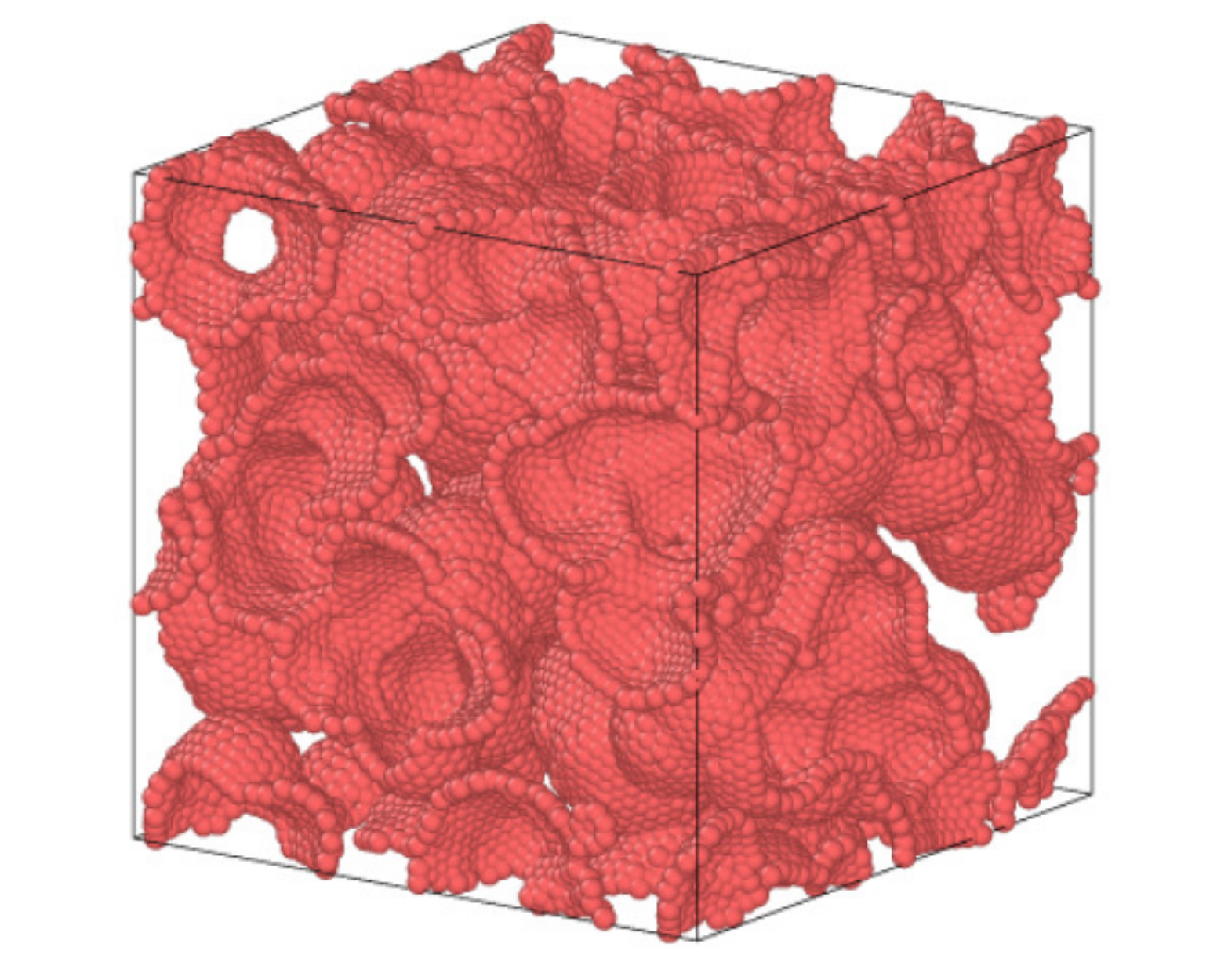}}
\subfigure[]{
\includegraphics[width=0.30\textwidth]{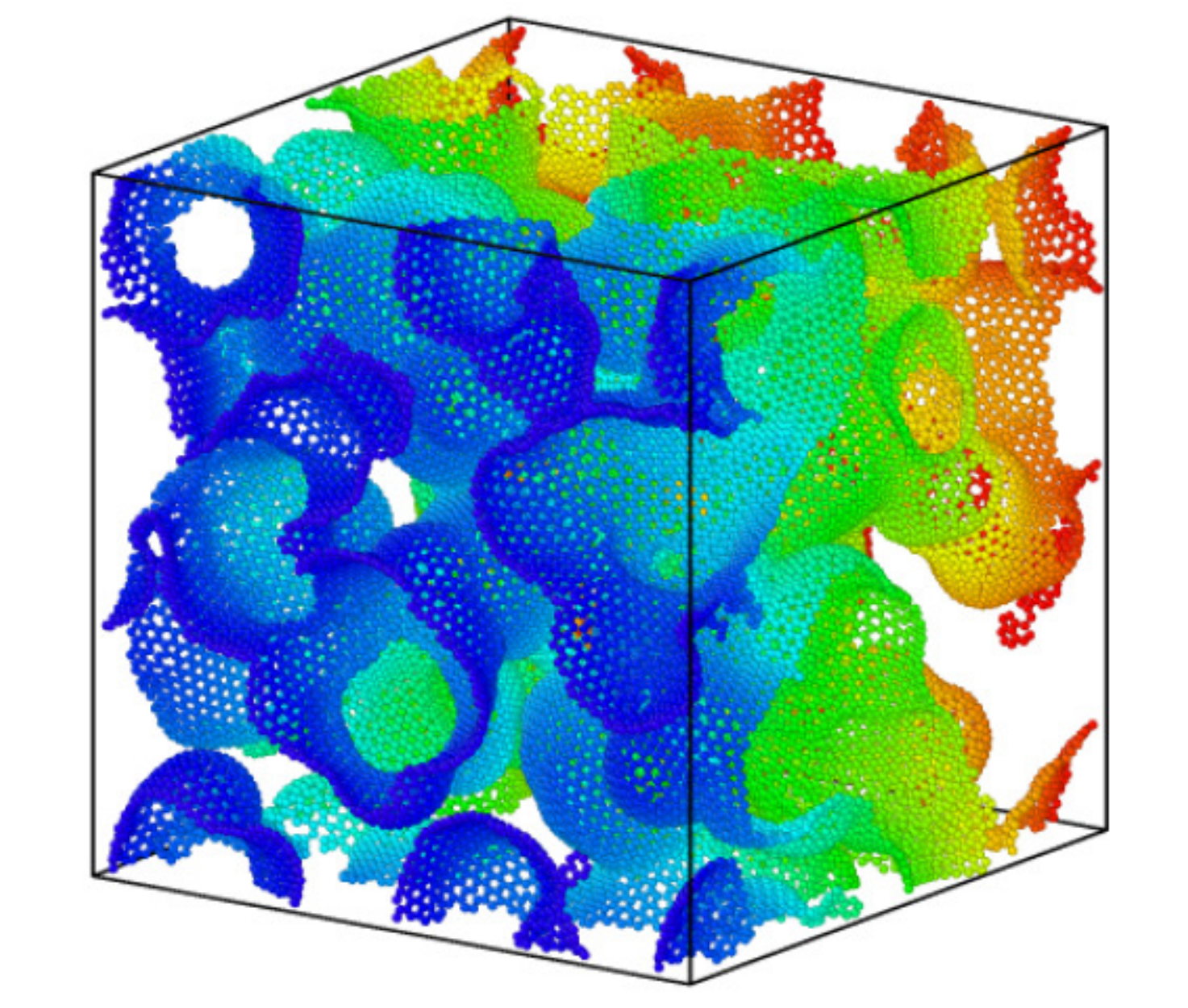}}
\caption{The step sequence for obtaining random foams. Panel a)
shows the initial condition in which the support particles are randomly arranged in a regular grid. Subsequently, the box is slowly expanded while the support particle positions are optimized (panel b). Foam particles are created on a regular grid and deleted if excessively close to the supporting ones (panel c). The particle positions are optimized and an attractive potential towards the supporting particles is switched-on during a molecular dynamics run (panel d). Finally, the particles that do not belong to the first layer are deleted to avoid multilayer structures (e). The LJ net is finally dualized by patterning the surface triangular tiling via a Voronoi procedure. A graphene-like topology eventually emerges (panel f). Color codes have been used for visualization purposes only and have no physical meaning.}
\label{fig:Procedure}
\end{figure*}

\begin{figure}
\centering
\includegraphics[width=0.4\textwidth]{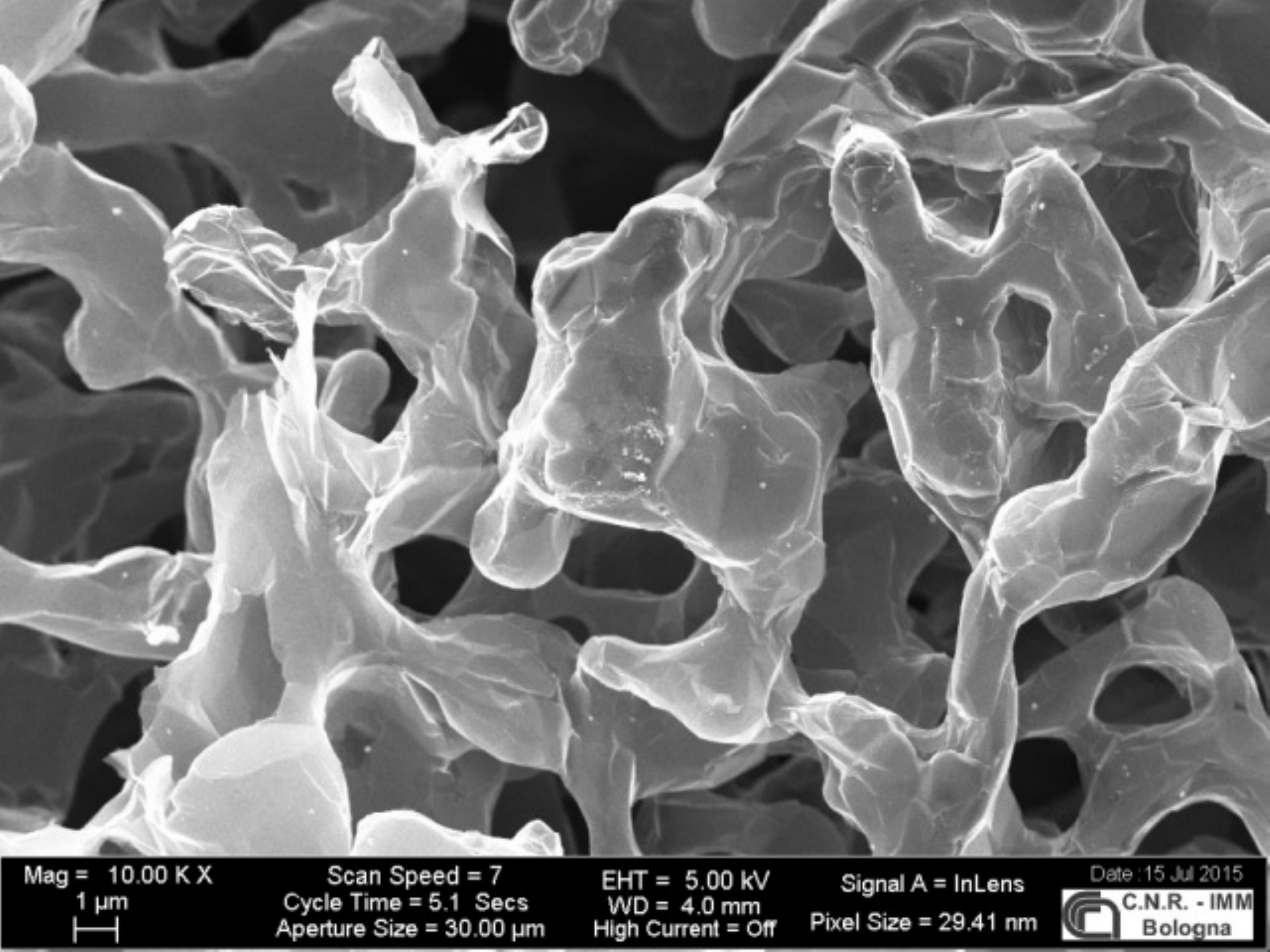}
\caption{Scanning electron microscopy image of a graphene random foam with a topology similar to those studied in this work (pore size is of course much larger, of the order of a few micrometers, with respect to that of our samples. (Courtesy of CNR-IMM Bologna, Italy.))}
\label{fig:FoamExp}
\end{figure}

This work is thus aimed at shedding some light on the mechanical and thermal properties of random graphene nanofoams. In particular, we present molecular dynamics (MD) simulations of random-pore foams under tension and compression  by modelling atomic interactions via reactive potentials. Several random-pore foams, characterized by different density and porosity, are produced using a tailored while reproducible recipe, which consists in preparing families of random networks to which graphene is attached. Mechanical properties are assessed by computing stress--strain curves, Young moduli, Poisson ratio, and specific toughness for each family of random foams. Furthermore, to assess the efficiency of our random-pore nanofoams as thermal insulators we report in this study the calculation of the effective phonon thermal conductivity by using the equilibrium Green-Kubo formalism.  

\begin{figure*}
\centering
\subfigure[A]{
\includegraphics[width=0.23\textwidth]{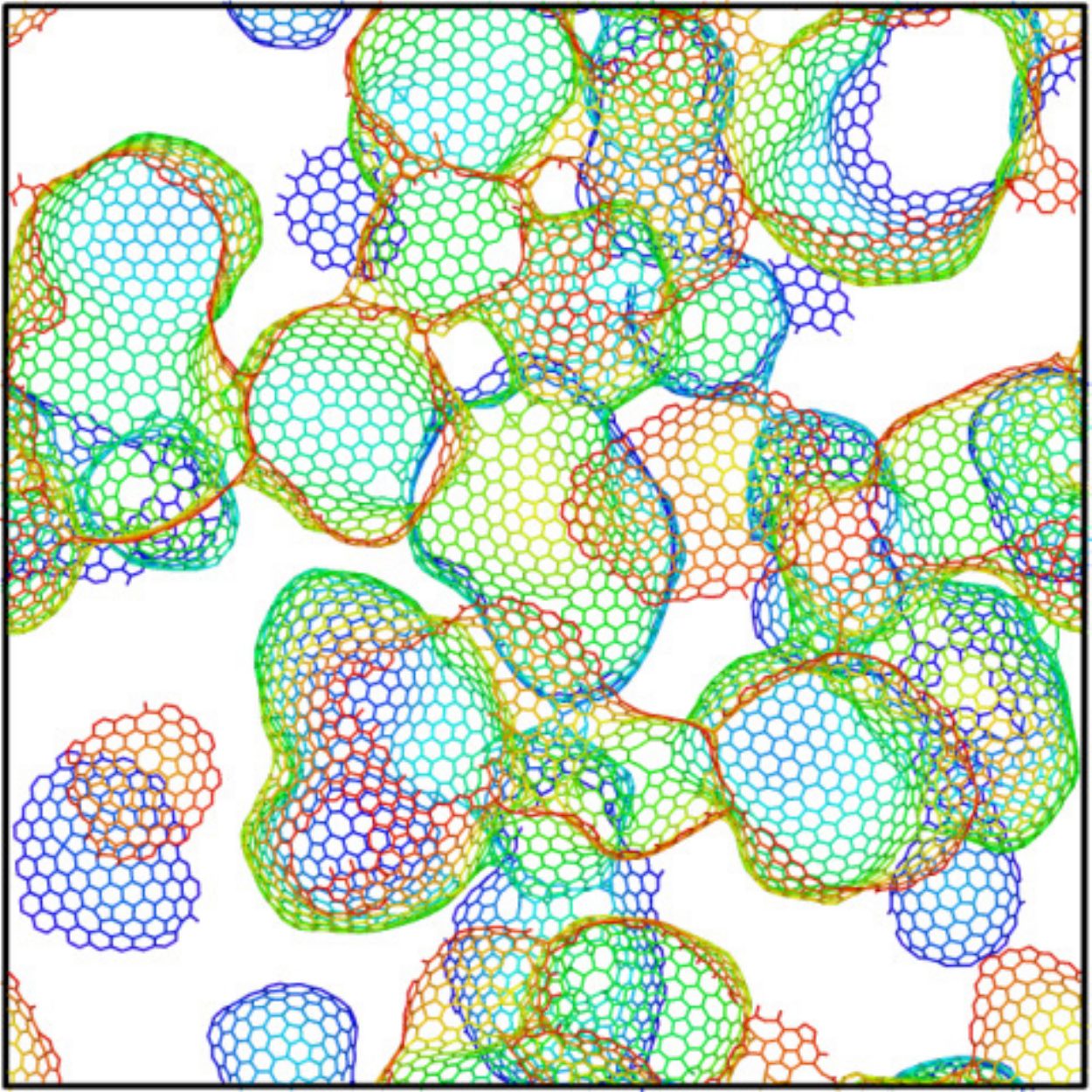}}
\subfigure[B]{
\includegraphics[width=0.23\textwidth]{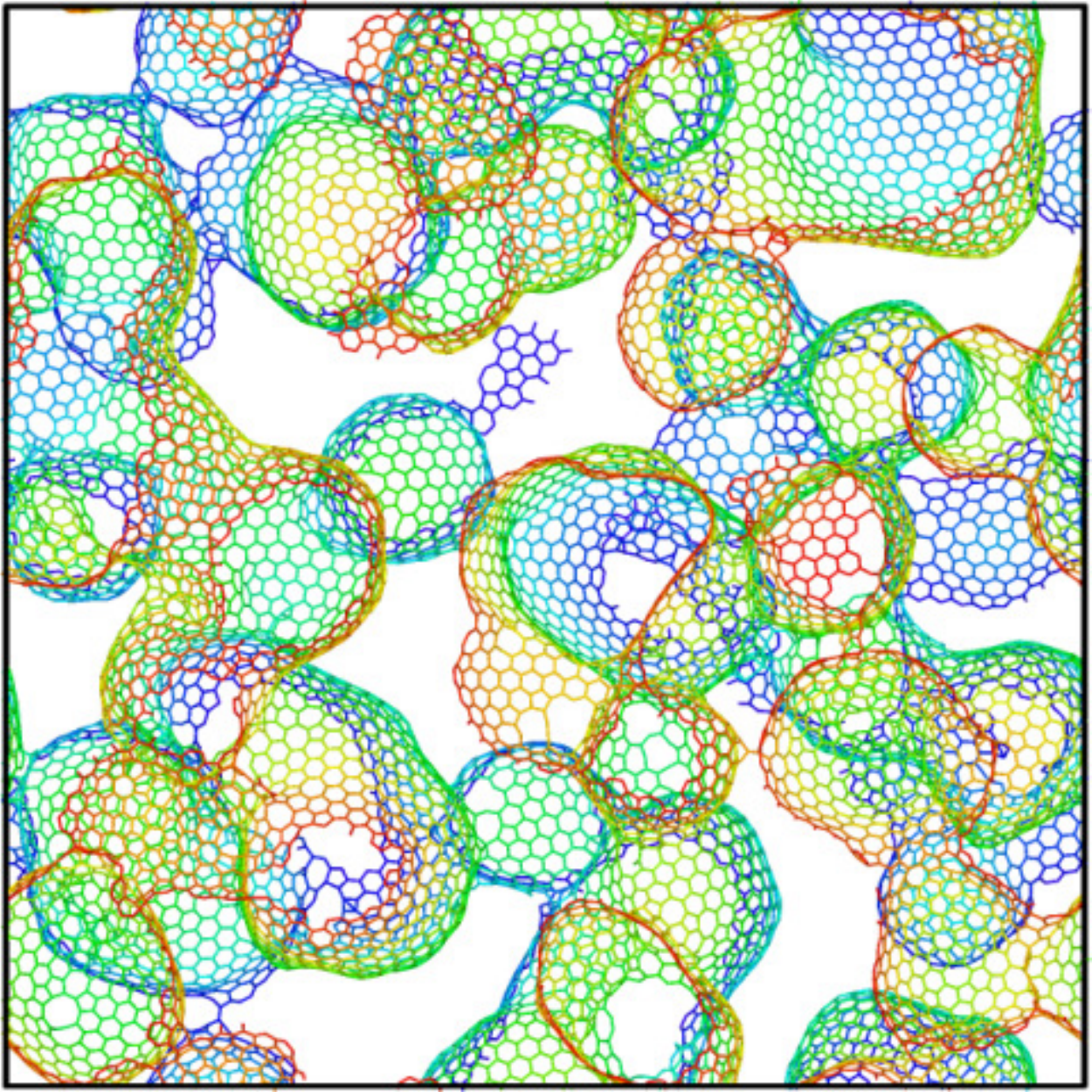}}
\subfigure[C]{
\includegraphics[width=0.23\textwidth]{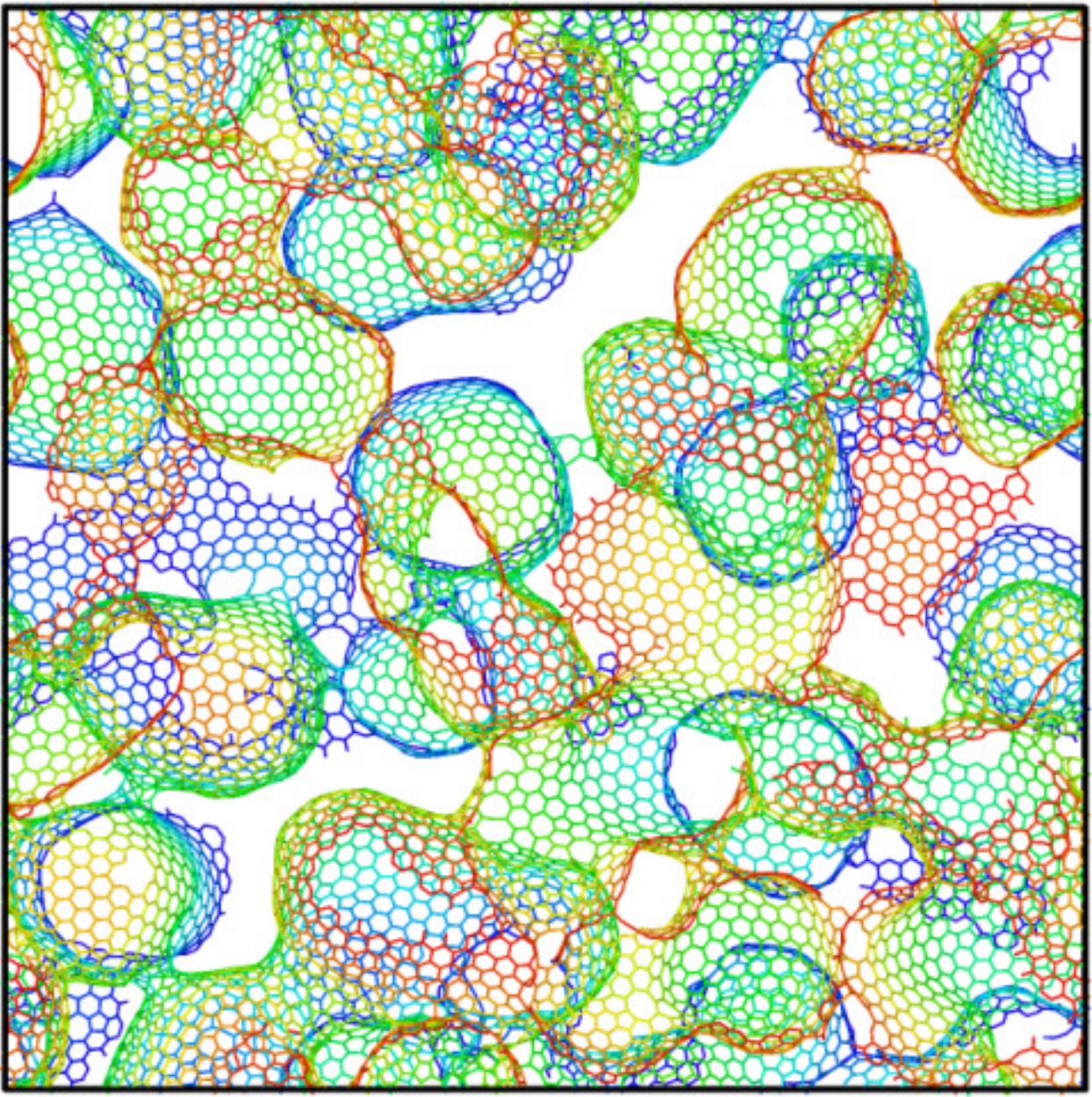}}
\subfigure[D]{
\includegraphics[width=0.23\textwidth]{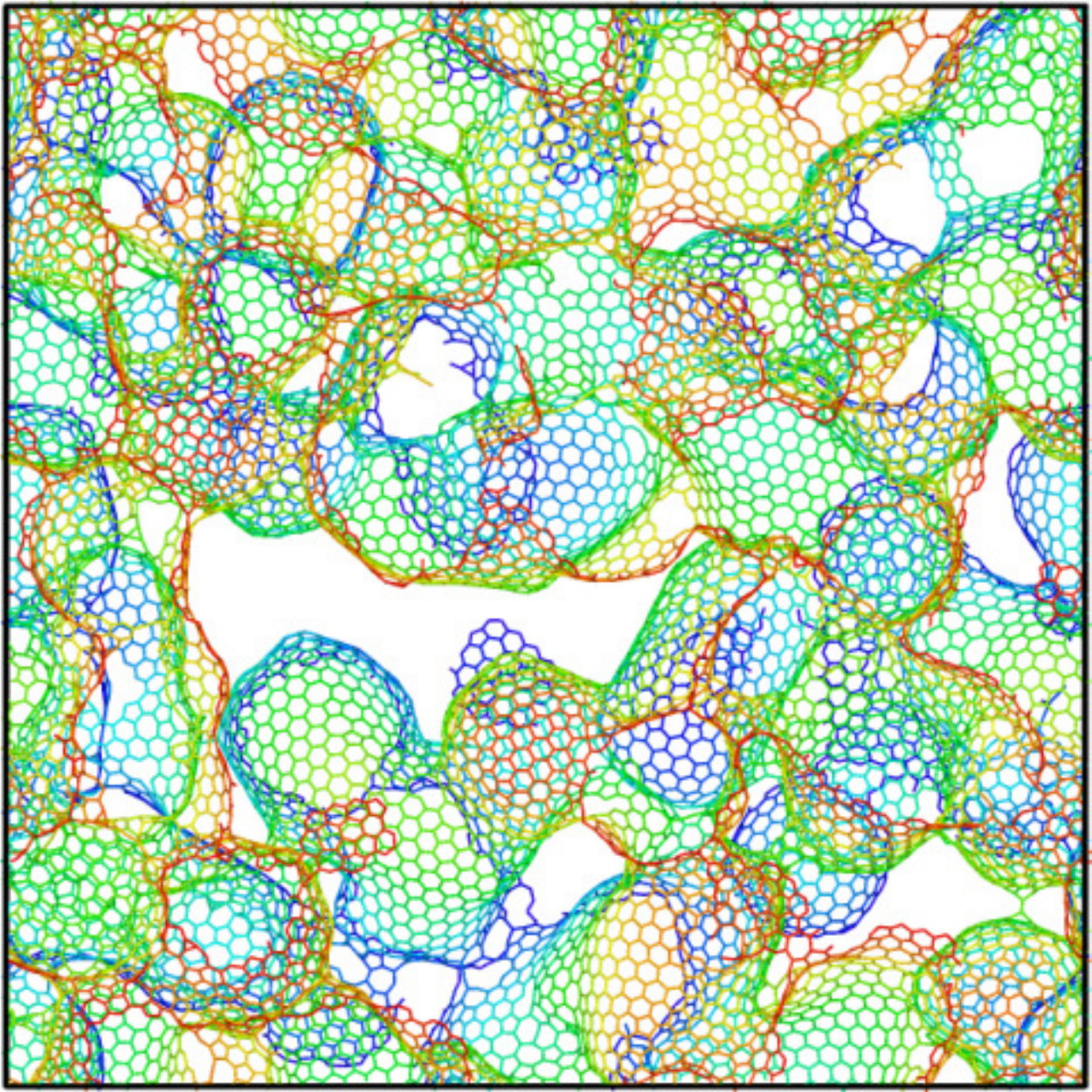}}
\caption{A $3.0$~nm slice of the unit cells for each of the four foam families with different porosity.}
\label{fig:Foams}
\end{figure*}

\section{Modeling graphene random foams}

To generate our families of graphene foams, we devise an approach composed basically of two steps ~\citep{Taioli2016,Pedrielli2017}: first, we generate a tessellation of the surface to be decorated with graphene by using triangles; second, we apply a Voronoi partitioning (dualization) of the triangulation points.

More in details, we start by filling the simulation unit cell with a random ensemble of particles interacting via a pair-wise Lennard-Jones (LJ) potential (Fig.~\ref{fig:Procedure}a). As a second step, the unit cell is slowly expanded to obtain random aggregation of particles (Fig.~\ref{fig:Procedure}b). In the third step we fill the simulation cell with a second type of particles arranged into a regular grid and characterized by different LJ parameters with respect to 
the previous ones (Fig.~\ref{fig:Procedure}c).
The first type of particles ({\it support particles}) acts as a framework to support the particles used in the triangulation ({\it foam particles}). The latter are deleted if they are too close (below $0.8$~nm) to the support particles in order to avoid convergence issues during molecular dynamics simulations. 
The fourth step consists in the optimization of the foam particle positions, performed by clamping down the framework degrees of freedom with a viscous damping force (Fig.~\ref{fig:Procedure}d). The particles found at a distance from the support particles larger than $0.32$~nm\ were deleted to obtain a smooth mono-layer structure (Fig.~\ref{fig:Procedure}e).  
As a last step, the Voronoi partition of the triangles tessellating the surface was performed to obtain pentagonal, hexagonal and heptagonal carbon rings (Fig.~\ref{fig:Procedure}f)~(for further details on this procedure see \citep{Taioli2016}). These configurations were finally annealed by MD using reactive potentials to optimize the carbon positions within the foams. 

By using this recipe, four families of carbon foams were produced (called A, B, C, and D, see Fig.~\ref{fig:Procedure}a, which are provided as xyz coordinate files with this submission). Each family is characterized by a different initial number of randomly-positioned support particles while, within each family, the only difference is the initial random distribution of the support points (Fig.~\ref{fig:Procedure}a).
During the whole procedure we impose periodic boundary conditions.

The LJ parameters used for the support (S) and foam (F) particles, respectively, are the following: 
$\epsilon_{SS}=100.0$~eV, $\sigma_{SS}=0.3$~nm, cutoff$_{SS}=0.5$~nm, $\epsilon_{FF}=0.1$~eV, $\sigma_{FF}=0.32$~nm, cutoff$_{FF}=0.23$~nm, $\epsilon_{SF}=10.0$~eV, $\sigma_{SFF}=1.0$~nm.
The starting cell side length is $6.0$~nm, expanded up to a length of $12.5$~nm from step a) to step b) of Fig.~\ref{fig:Procedure}. 

\begin{table}[t]
\centering
\small
\begin{tabular}{cccccc}
\toprule
                & Average   &    Average      &  Carbon atoms with      &    \\ 
    Foam type   & density   &    pore size     & 3-coordination  &   \\ 
                & (g/cm$^3$)               &   (nm)  &  (\%)     &   \\
\midrule
A & 0.55 & 	2.23  &    97.3 &    \\ 				
B & 0.68 &   1.89 &    95.9  &  \\ 
C & 0.78 &  1.68 &   	93.0  &   \\ 
D & 0.83 &  1.56 &    93.1  &   \\ 				
\midrule

\bottomrule
\end{tabular}
\caption{Parameters characterizing the four foam families investigated in this work.}
\label{tab:Table 1}
\end{table}
These parameters were chosen in such a way that the typical distance between the support particles was smaller than the equilibrium length between the foam and the support particles. In this way, the support particle surface is smooth, being obtained by several atoms lying nearby.

The topology of the nanofoams studied here is inspired by the graphene foams grown on nickel scaffolds (Fig. \ref{fig:FoamExp}). However, we note that the pore size in the experimentally synthesized samples is larger than in our computational models. The way in which our graphene random foams are built is substantially different from that one presented in Ref. \citep{Qin2017}, 
where three-dimensional graphene assemblies were synthesized by starting from randomly distributed and oriented rectangular graphene flakes and spherical inclusions, and by repeating NPT-NVT cycles to obtain condensed graphene foams.

\section{Characterization of graphene foams}

Five different samples for each of the four families were prepared by varying the initial distribution of the support particles. In Fig. \ref{fig:Foams} we report representative models ($3.0$~nm slices) for each of these foam families. From A to D the foams present a decreasing average pore size, and an increasing mass density.

The geometrical analysis of the graphene porous foams and of their voids was carried out using the simulation code Zeo++\citep{Williems2012}. In particular, we characterize our prepared structures using the Pore Size Distribution (PSD) function, which can be experimentally obtained by adsorption/desorption measurements. PSD analysis delivers a quantitative description of the range of pore sizes present in a given sample. 

Moreover, we perform a coordination analysis to find possible signature of under- or over-coordination of the carbon atoms usually forming a network of sp$^2$ hybrid bonds. The computed quantities are reported in Tab~\ref{tab:Table 1}. 

\begin{figure}  
\includegraphics[width=0.45\textwidth]{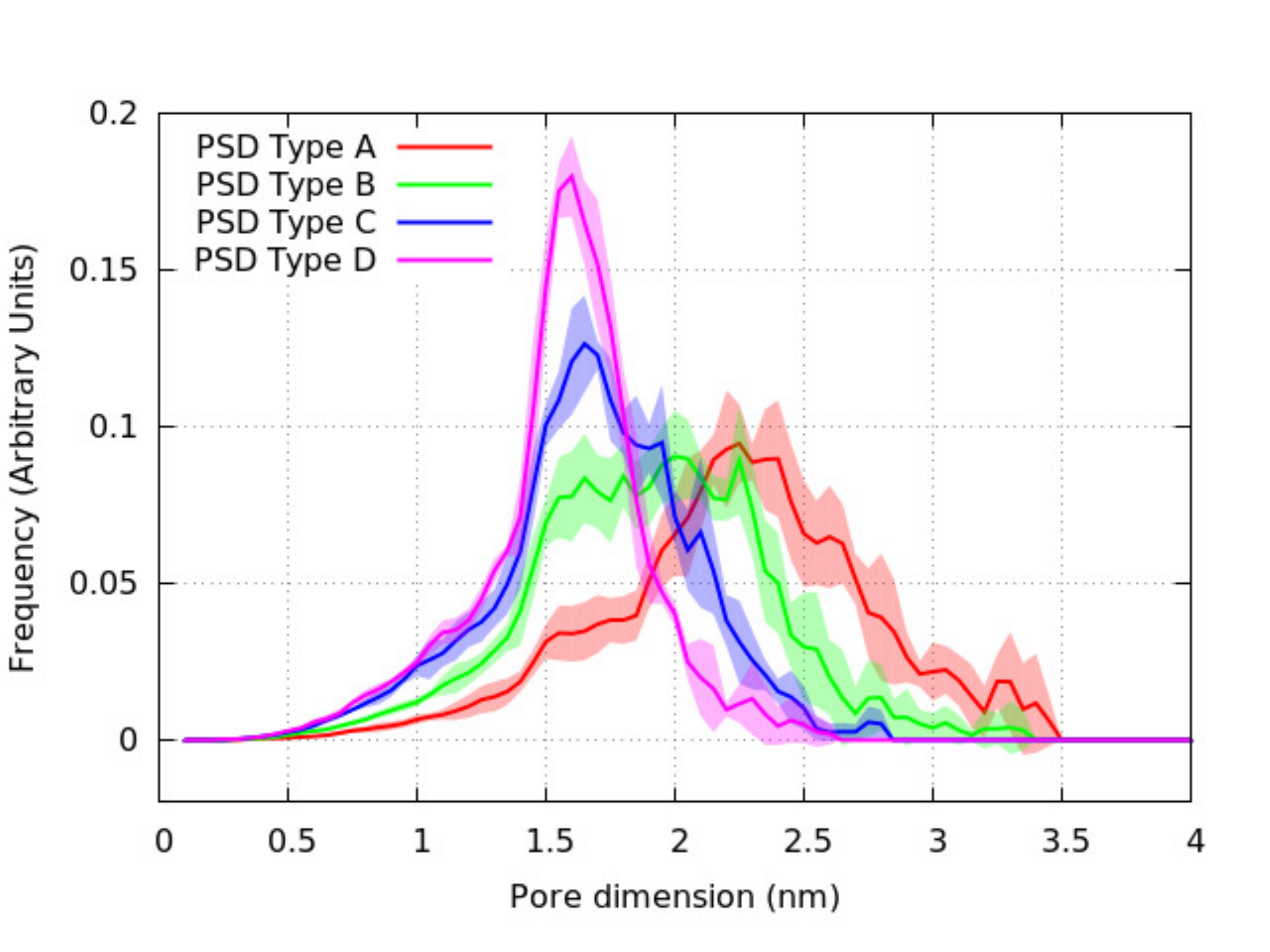}
\caption{Pore Size Distribution of the four families of random nanofoams. The average pore size of the considered foam families is a decreasing function of mass density.}
\label{fig:PoreSize}
\end{figure}

\begin{comment}
The PSDs can be obtained as an overlap of two peaks: the first, centered at $1.7$~nm, is related to the internal space of the random foam; the second, variable between $2.3$~nm towards $1.7$~nm, is attributed to the external space of the foam.
\end{comment}

\begin{figure} 
\includegraphics[width=0.45\textwidth]{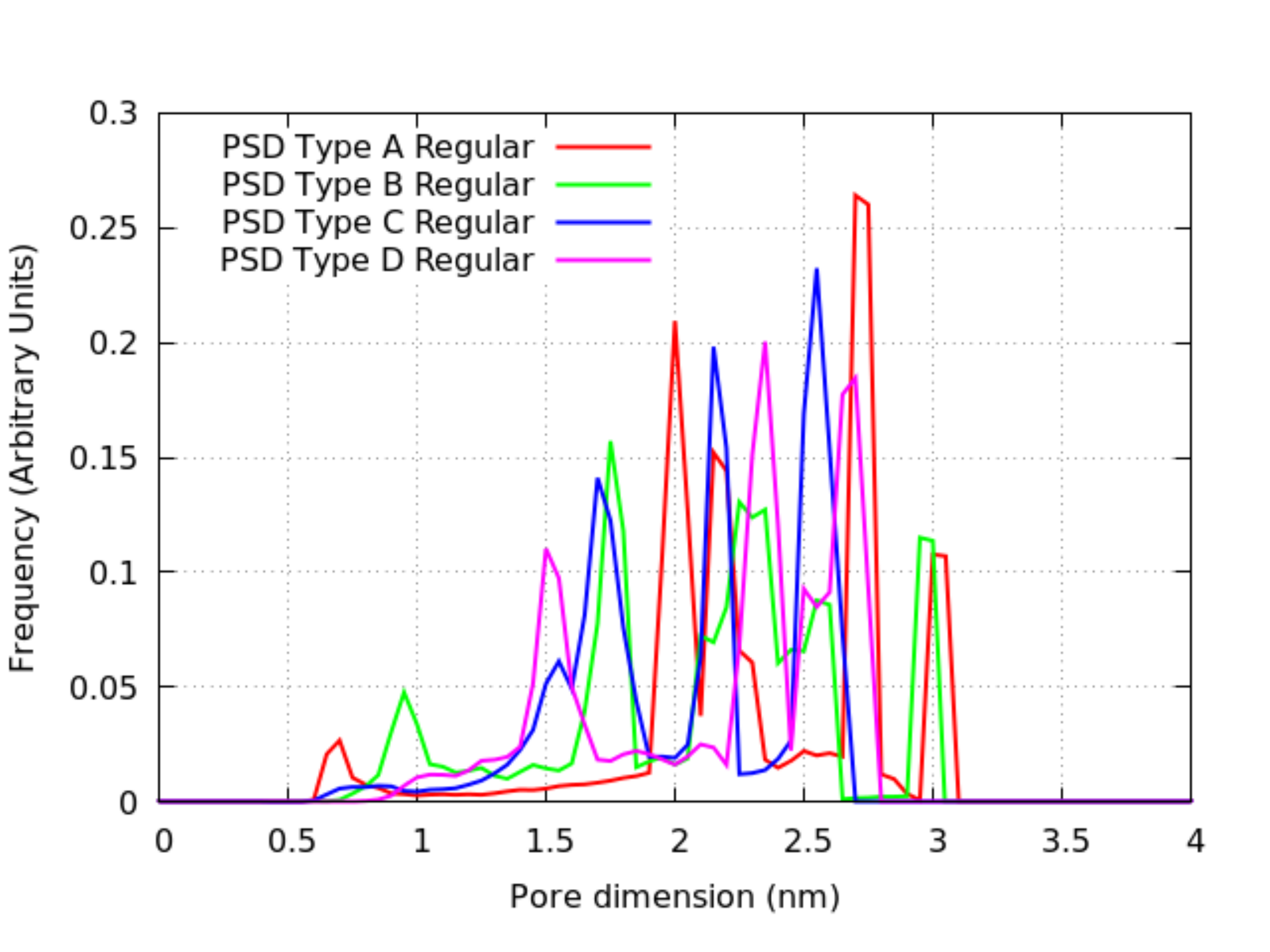}
\caption{Pore Size Distribution of the four types of regular nanofoams, such as those presented in Ref. \cite{Pedrielli2017}. The average pore size is $2$~nm, comparable with the random foam studied here, in particular with foams type B and C reported in Fig. \ref{fig:PoreSize}.}
\label{fig:PoreSizeRegular}
\end{figure}

\begin{figure}  
\includegraphics[width=0.45\textwidth]{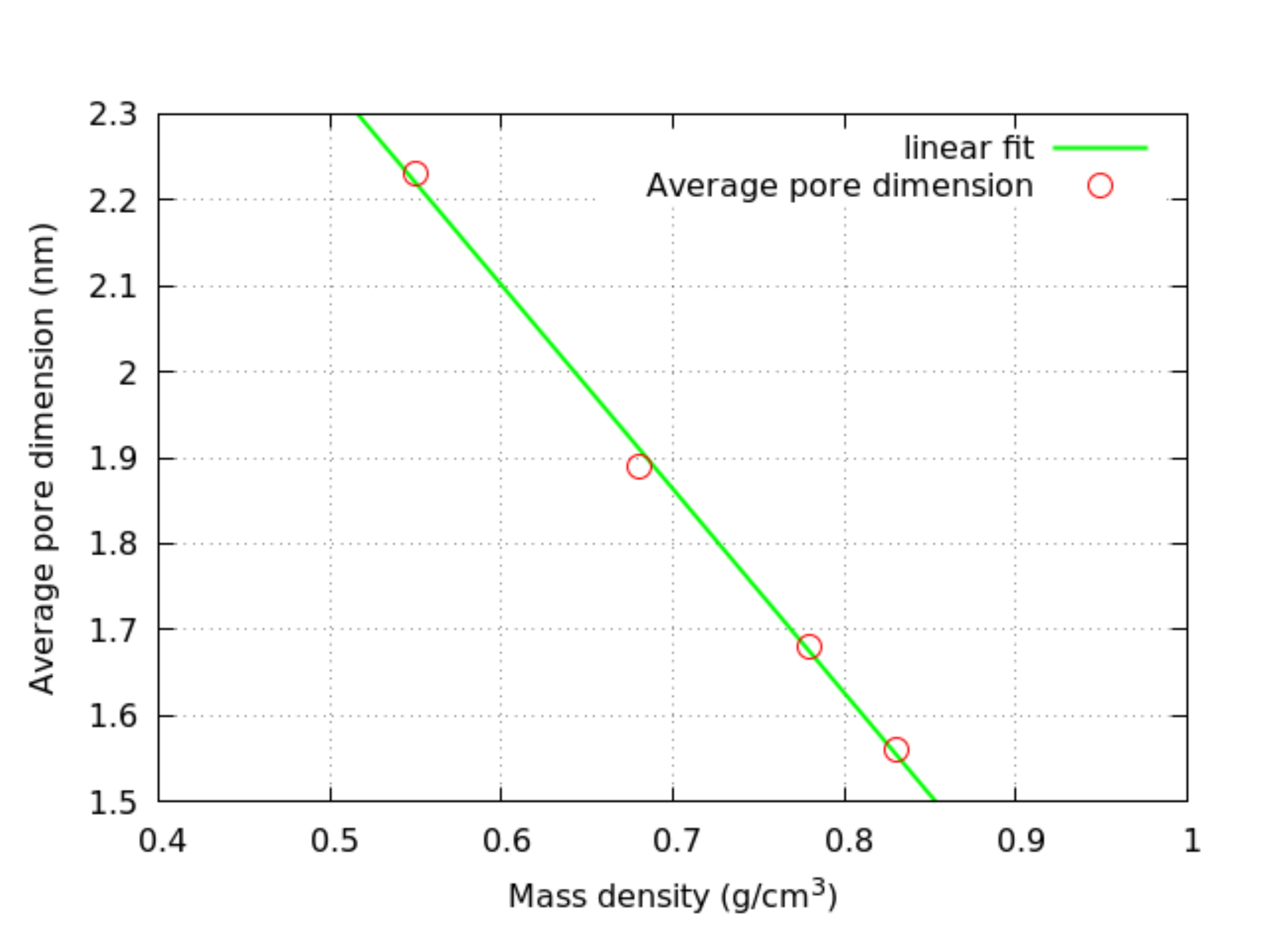}
\caption{Relation between mass density and average pore dimension for the considered foam families. }
\label{fig:PoreMass}
\end{figure}

The averaged PSDs for all our graphene foam families are reported in Fig.~\ref{fig:PoreSize} (continuous lines), showing the standard deviation within each group as a colored shaded area. By comparing these PSDs with those obtained in the case of regular pore foams \citep{Pedrielli2017}, reported in Fig. \ref{fig:PoreSizeRegular}, we notice that the random foams present similar average pore dimension and similar mass density. Indeed, the random foam PSDs are characterized by a maximum, representing the most likely pore size in each case, decreasing from $2.3$~nm to $1.7$~nm from A to D foam type. The average values of the pore size are reported in Tab. \ref{tab:Table 1}.
These values compare reasonably well with those reported for regular foams, where PSD peaks at about $2$~nm (see Fig. \ref{fig:PoreSizeRegular}). 
\begin{comment}
We notice that the PSDs of random foams (see  Fig.~\ref{fig:PoreSize}) can be rationalized as overlap of two contributions: a smaller pore located in the interior of the foam, identified by the peak at $1.7$~nm common to all families; and a second peak decreasing from $2.3$~nm towards lower values with increasing foam density.
Thus, due to this double peak feature, our structures can represent a rather faithful model of foams in which the main average dimension of the internal pore results essentially unchanged under compression, whereas the interstitial space decreases progressively upon compression having pores with lower average dimensions.
\end{comment}
Finally, while regular foams present mass densities in the range $0.6-0.7$~g cm${^{-3}}$, our families of random foams have mass densities in the range $0.5-0.8$~g cm${^{-3}}$ (see Tab. \ref{tab:Table 1}).
We notice that the an almost linear relation (coefficient of determination R$^2=0.98$) with negative slope relates the mass density and the average pore dimension of the four random foam families, as reported in Fig. \ref{fig:PoreMass}.

\section{Computational methods}

To perform molecular dynamics simulations, carbon-carbon interatomic forces were modeled using the AIREBO potential \citep{Stuart2000}. 
To find the minimum energy structures with respect to defect positions, the samples were annealed at $3500$~K, equilibrated at this temperature for $100$~ps, and eventually cooled down to $700$~K in $100$~ps using a viscous damping force. The annealing was performed using the standard value for the cutoff parameter for the REBO part of the potential and performed within the microcanonical ensemble (NVE).  
For the simulations of compressive and tensile regimes, all samples were equilibrated at zero pressure and at the temperature of $1$~K using the Nosé–Hoover barostat and thermostat.
Furthermore, the adaptive cut-off parameter of the potential was set to $0.2$~nm to better describe the near-fracture regime \citep{Shenderova2000}. The equations of motion were integrated via the velocity-Verlet algorithm with time step of $1$~fs. Mechanical properties were assessed in the isobaric-isothermal ensemble (NPT), adding a drag term to smooth out the pressure oscillations. 

A uni-axial tensile strain was applied up to sample fracture in each case. The strain parallel to the direction of deformation is defined as
\begin{equation}
\varepsilon = \frac{L-L_0}{L} = \frac{\Delta L}{L}
\end{equation} 
where $L_0$ and $L$ are the starting and current length of the 
sample in the direction of loading. To determine the stress, the pressure stress tensor components in response to the
external deformation are computed as \citep{Thompson2009}
\begin{equation}\label{pressure}
P_{ij} = \frac{\sum_k^N{m_k v_{k_i} v_{k_j}}}{V}+ \frac{\sum_k^N{r_{k_i} f_{k_j}}}{V}
\end{equation} 
where $i$ and $j$ label the coordinates $x$, $y$, $z$; $k$ runs over the
atoms; $m_k$ and $v_k$ are the mass and velocity of $k$-th atom; $r_{k_i}$ is the position of $k$-th atom; $f_{k_j}$ is the $j$-th component of the total force on the $k$-th atom due to the other atoms; and, finally, $V$ is the volume of the simulation box.

The pressure in Eq. \ref{pressure} includes both kinetic energy (temperature) and virial terms. Notice that the forces appearing in Eq. \ref{pressure} are the sum of the pairwise, angle, dihedral, improper and long-range contributions. The computed stress is the true stress because the pressure is measured with respect to the instantaneous section area of the samples. The uni-axial compressive strain was applied up to reaching $35$~\% total strain. The applied strain rate is chosen equal to $0.001$~ps$^{-1}$, that we tested appropriately for ensuing convergence in the case of regular nanofoams \citep{Pedrielli2017}. Stress and strain were saved every $1000$ time steps.
Foams are rather rigid materials before packing, so one needs to apply large loads, of the order of a few GPa, to obtain large deformations in comparison to other materials, such as concrete or steel (having strains of 0.01\% already at pressures of few hundreds MPa). However, to reach the highest level of deformation (up to 35\%) in the foams studied in this paper one should resort to using diamond anvil cells or small samples in order to achieve pressures of a few GPa.

The stress--strain curve was computed at $1$~K, since molecular dynamics is usually computationally faster than minimization procedures. The same approach has been previously adopted by other groups dealing with similar problems (see e.g. \cite{Wu2013}). 
While thermal fluctuations of the order of a few K, thus higher than the absolute value of the thermostat temperature, are found during MD runs, they do not significantly affect numerical MD simulations. Indeed, we demonstrated in a previous work [1] that the contribution of the kinetic energy to the pressure tensor at a few K is approximately 2\% of the total. Thus, a small kinetic contribution due to using low temperature MD does not prevent our simulations from reaching and overpassing local minima. The use of low-temperature MD simulations was chosen instead of standard minimization procedures because stress, for example due to compressive load, can be more effectively and more continuously redistributed within the entire structure during the dynamics by applying a deformation rate (providing this rate delivers converged results with respect to its value) instead of using a sequence of deformation-minimization steps. Indeed, under compressive strain the temperature is likely to increase: coupling the system with a thermostat leaks away this excess of thermal energy and allows for a minimisation of the structures by using MD.

The observables that we calculate to characterize the mechanical properties of the nanofoams are the Young modulus, fracture stress and fracture strain. The toughness is also evaluated as the area under the stress--strain curve up to the fracture stress. Indeed, the samples have no plastic deformation but several sequential fractures.
Stress--strain characteristics of carbon random nanofoams present a linear behaviour at low strain. Thus, the Young modulus is obtained as the tangent at zero strain. 

We also performed the calculation of the Poisson ratio $\nu$, defined as the negative ratio between the transverse $\varepsilon_{\mathrm{T}}$ and the longitudinal deformations $\varepsilon_{\mathrm{L}}$:
\begin{equation}
\nu =- \frac{\varepsilon_{\mathrm{T}}}{\varepsilon_{\mathrm{L}}}
\end{equation} 

Here we extend the concept of Poisson ratio to deformations beyond the linear regime, and use it to quantify the lateral deformation of the material. A similar extension is done for the Young modulus.

Phonon thermal conductivity was assessed using the  equilibrium Green-Kubo approach \citep{Green1954, Kubo1957} for it is less sensitive to the simulation box dimension than non-equilibrium MD methods \citep{Sellan2010}.
To this aim, first the atomic positions were relaxed and equilibrated at $300$~K using the Berendsen thermostat method (NVT ensemble).

Then, in the NVE ensemble, the equilibrium thermal conductivity $k$ according to the Green-Kubo formalism, can be calculated as follows:

\begin{equation}\label{cond}
k = \frac{V}{3K_{\text B}T^2} \int_0^{\infty}{\langle \vec{J}(0) \cdot  \vec{J}(t)\rangle}dt
\end{equation}

where $V$ is the volume of the simulation cell, $t$ is the correlation time, $K_{\text B}$ is the Boltzmann constant, $\vec{r}$ identifies the particle positions. The heat current $\vec J$, appearing in Eq. \ref{cond}, is defined by: 

\begin{equation}\label{energy}
\vec{J} = \frac{1}{V} \Big( \sum_{i} E_i \vec{v}_i + \frac{1}{2} \sum_{i<j}(\vec{F}_{ij} \cdot (\vec{v}_i + \vec{v}_j ) \vec{r}_{ij}) \Big)
\end{equation} 

where $\vec{v}$ is the velocity of a particle, $\vec{r}_{ij}$ and
$\vec{F}_{ij}$  are the distance and force between the particles $i$ and $j$, and $E_i$ is the total energy per atom. The first term in the right hand side corresponds to convection, while the second term to conduction. The integrand in the expression for thermal conductivity is the heat current auto-correlation function (HCACF). 
To get a proper sampling of the phase space multiple runs are required with different initial conditions. Simulations to obtain MD trajectories to perform accurate ensemble averages were performed over a time span of $500$~ps, using a step of $0.5$~fs. HCACF has been computed by dividing the total time of computation into $250$~fs beads and by performing the integral in Eq. \ref{cond} by sampling every $5$~fs. Finally, we average over all the beads.

The thermal conductivity was calculated by using a version of the Tersoff potential \citep{Lindsay2010} optimized to reproduce accurately the experimental phonon dispersion curves and the thermal properties of carbon structures, such as graphene and graphite.

Molecular dynamics simulations were carried out using LAMMPS \citep{Plimpton1995}. Atomic configurations were visualized by using the OVITO package \citep{Stukowski2010} or VMD \citep{Humphrey1996}. 

\section{Results and discussion}

\subsection{Tension}

\begin{figure}[hbt]
\centering
\includegraphics[width=0.5\textwidth]{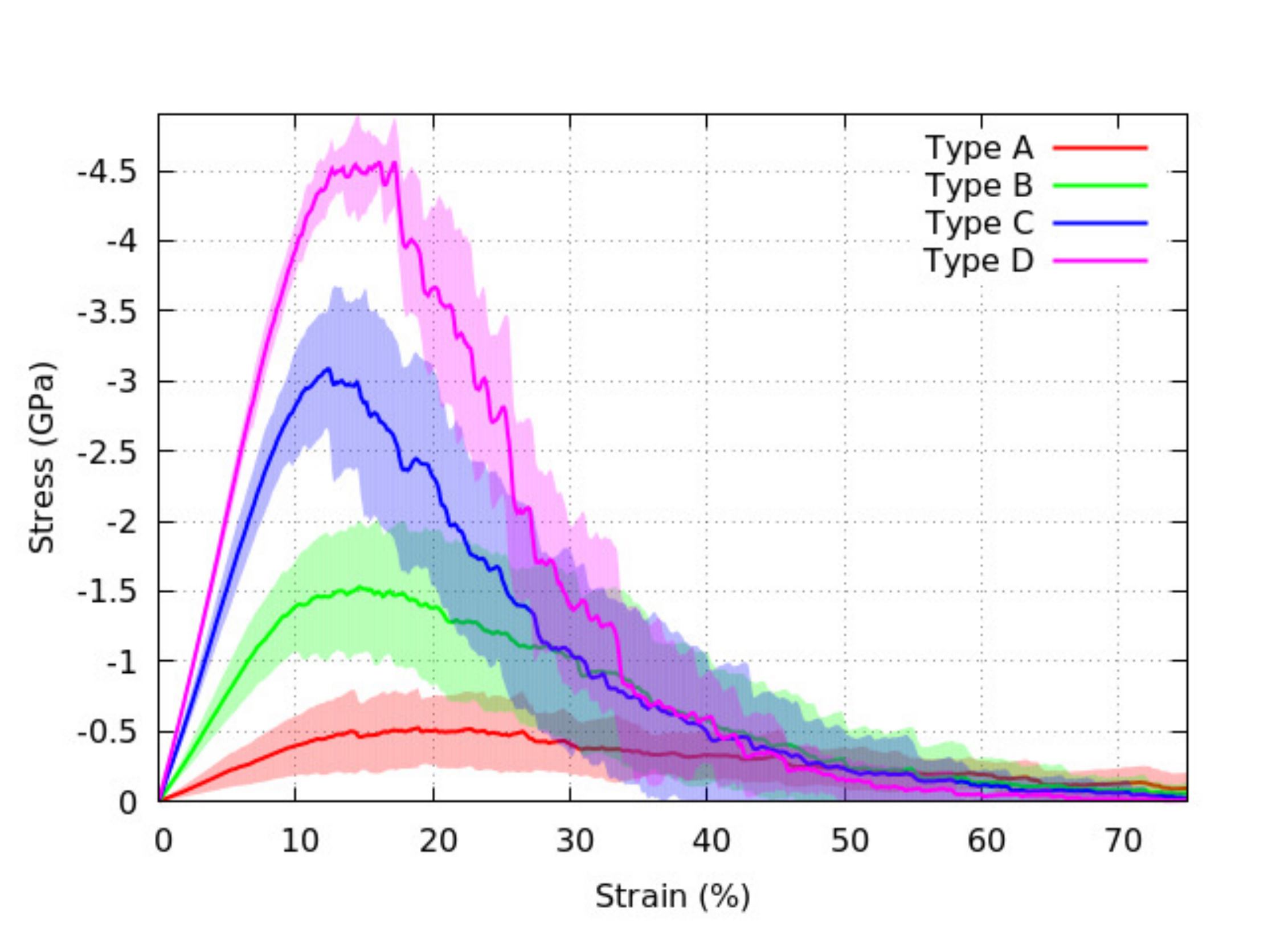}
\caption{Stress--strain curves of the four graphene foam families under uni-axial tension along with the standard deviation for each family, reproduced by a colored shaded area on the top of the relevant curve.}
\label{fig:Tension}
\end{figure}

\begin{figure}[hbt]
\centering
\includegraphics[width=0.5\textwidth]{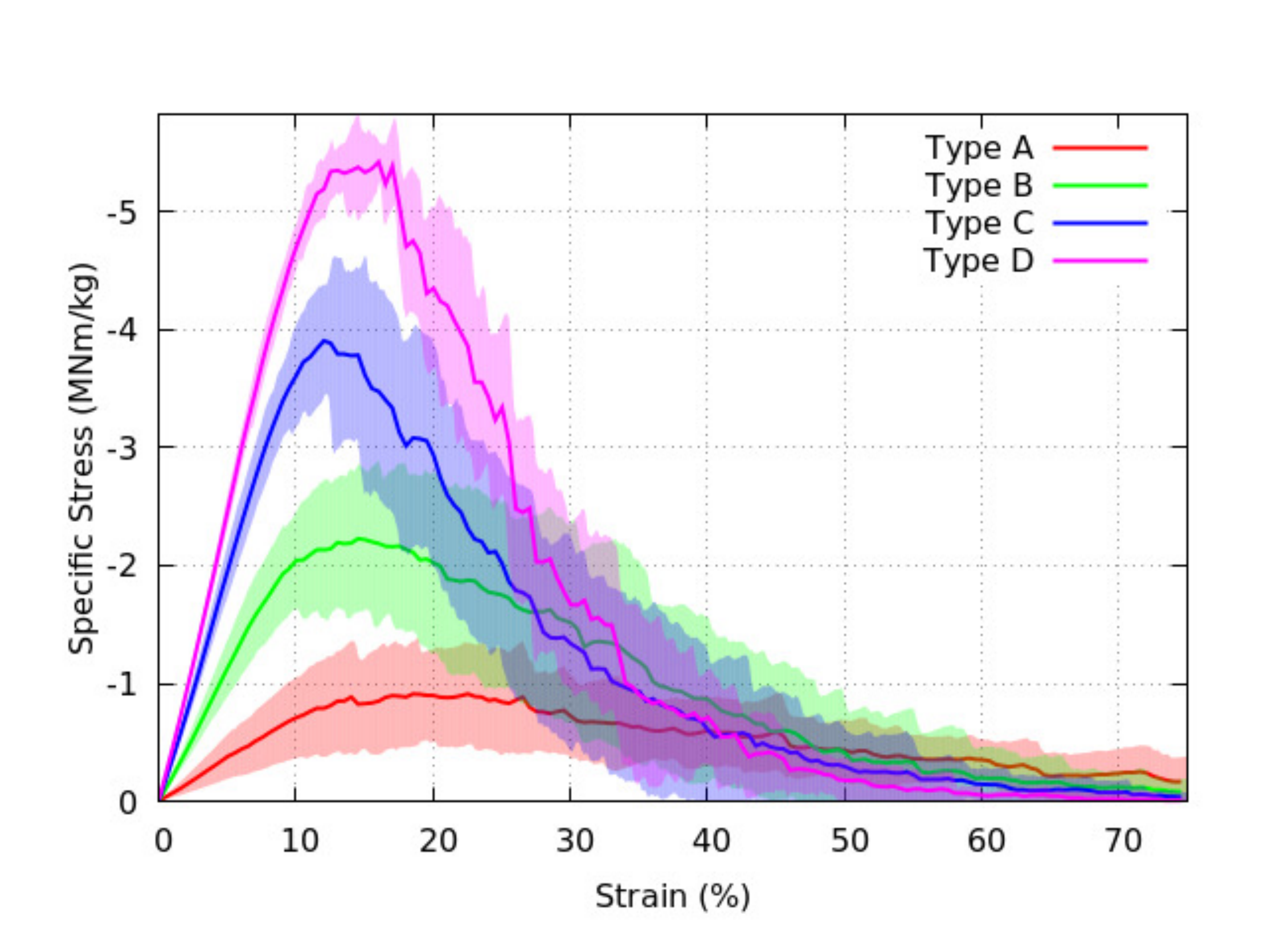}
\caption{Stress--strain curves of the four foam families weighted by the sample mass density under uni-axial tension along with the standard deviation for each family, reproduced by a colored shaded area on the top of the relevant curve.}
\label{fig:TensionSpec}
\end{figure}

In Fig.~\ref{fig:Tension}, we report the stress--strain characteristics for the four foam families investigated in this work, while in Fig.~\ref{fig:TensionSpec} the stress--strain curves are normalized with respect to the mass density. 

The stress--strain curves show a typical elastic behavior for small deformations up to the tensile strength, followed by a decreasing tail corresponding to the sample fracture. We notice that the stress--strain characteristics of graphene foams do not present a region that can be associated to a plastic deformation. Indeed, these 3D graphene structures are essentially brittle, presenting a comparable fracture strain with a corresponding stress specific of the family. Notably, the same behaviour is found in the mass density weighted stress--strain curves (see Fig.~\ref{fig:TensionSpec}).
This finding tells us that the different mechanical performances of the four foam families are due basically to features other than mass density, such as the pore size distribution or the connectivity. Indeed, if the mass density were the factor most critically affecting the foam mechanical properties, then the normalized stress-strain curves should overlap.

Moreover, we report in Tab.~\ref{tab:Table 2} the Young modulus and the tensile strength of the four foam families, and in Tab.~\ref{tab:Table 2b} the specific modulus and the specific strength (values per mass density). Furthermore, in the fifth column of Tab.~\ref{tab:Table 2b} we show the specific toughness, calculated as the total area below the stress--strain curves of Fig.~\ref{fig:TensionSpec} up to the fracture strain.
Specific Young modulus and tensile strength of random foams can be compared with those previously calculated for carbon nanotruss networks, studied in Ref. \citep{Pedrielli2017}. 
For nanotruss network, at $5$\% to $8$\% strain 
the stress is in the range $90-130$~MNm~kg$^{-1}$, while for random foams at the same strain the values are in the range $3.9-36.6$~MNm~kg$^{-1}$. This makes clear that regular foams are mechanically stiffer than the random ones here studied. Graphene random foams can be also compared to 3D graphene assemblies reported in Ref. \citep{Qin2017}. Graphene assemblies have a specific Young modulus of $7.65$ MNm kg$^{-1}$ (mass density: $0.366$ g/cm$^{3}$, Young modulus: $2.8$~GPa), which compares with the lowest mass density foam family studied here (see Tab. \ref{tab:Table 2}). At variance, 
the specific strength of $7.4$ MNm kg$^{-1}$ found in graphene assemblies is only sligthly higher than in our random foams, mainly due to the higher connectivity of graphene sheets composing the assemblies. For completeness, we notice that the mechanical tests reported in Ref. \citep{Qin2017} have been performed at a temperature of $300$~K, while our simulations are performed at $1$~K. Movies of our foams under tensile load are provided with this submission.

\begin{table*}
\centering
\small
\begin{tabular}{cccccccc}
\toprule
             & Young     & Standard  & Tensile   & Standard  & Fracture  \\ 
   Foam type   & modulus  & deviation  &  strength & deviation &  strain     \\ 
             &(GPa)      & (GPa)  &   (GPa)    & (GPa) &     (\%)    \\
\midrule
A & 3.9  & 2.1 & 0.5 &  0.3    &  19 \\ 
B & 14.0 & 3.1 & 1.5 &  0.5    &  15 \\ 
C & 27.3 &  3.7 & 3.1 &  0.4   &  12 \\ 
D & 36.6 &  1.9 & 4.6 &  0.3   &  13  \\ 
\bottomrule
\end{tabular}
\caption{Young modulus (1$^{st}$ column) and its standard deviation (2$^{nd}$ column), tensile strength (3$^{rd}$ column) and its standard deviation (4$^{th}$ column), and fracture strain (5$^{th}$ column) of
the four families of random foams under tension. Specific toughness is calculated as the area below the stress--strain curve up to fracture strain per mass density.}
\label{tab:Table 2}
\end{table*}

\begin{table*}
\centering
\small
\begin{tabular}{ccccccccc}
\toprule
           &    Specific  & Standard & Specific  & Standard &  Specific\\ 
 Foam type    &  modulus  & deviation & strength  & deviation & toughness\\ 
           &   (MNm kg$^{-1}$) & (MNm kg$^{-1}$) & (MNm kg$^{-1}$) & (MNm kg$^{-1}$)&    (MJ~kg$^{-1}$) \\
\midrule
A & 7.1 & 3.8 & 0.9   &0.5&  0.1 \\ 
B & 20.6 & 4.6 &  2.2 &1.5&  0.2 \\ 
C & 35.0 & 4.7 & 3.9  &3.1&  0.3 \\ 
D & 44.1 & 2.3 & 5.4  &4.6&  0.4 \\ 
\bottomrule
\end{tabular}
\caption{Specific modulus (1$^{st}$ column) its standard deviation (2$^{nd}$ column), specific strength (3$^{rd}$ column) and its standard deviation (4$^{th}$ column), and  specific toughness (5$^{th}$ column) of
the four families of random foams under tensile strain. Specific toughness is calculated as the area below the stress--strain curve up to fracture strain per mass density.}
\label{tab:Table 2b}
\end{table*}

\subsection{Compression}

\begin{figure}[hbt] 
\centering
\includegraphics[width=0.5\textwidth]{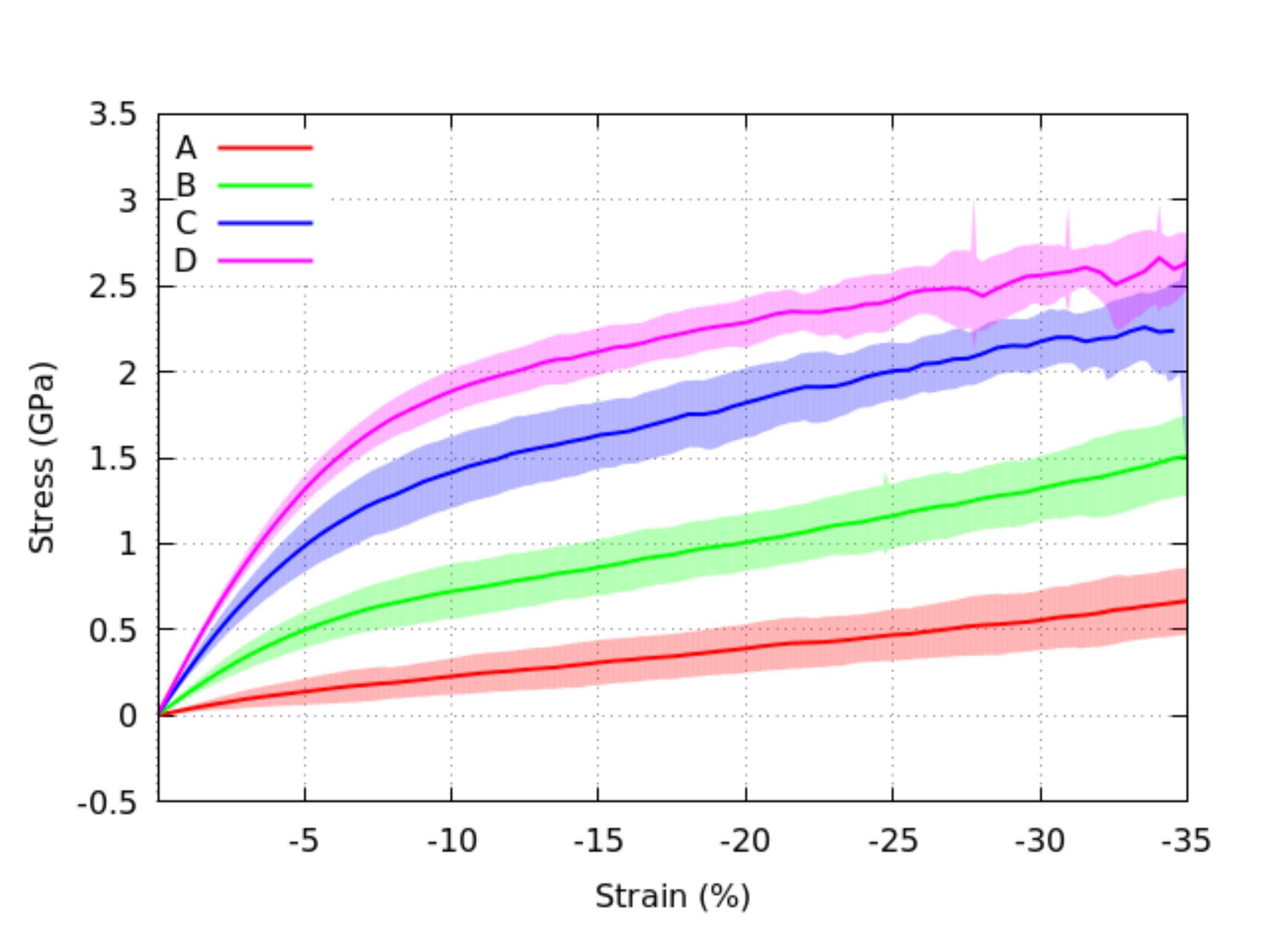}
\caption{Stress--strain curves of the four random foam families under uni-axial compression up to $35$\% strain. Shaded areas represent standard deviation within each foam family.}
\label{fig:Comp}
\end{figure}

In this section we present the results obtained for our samples under compressive load. In Fig.~\ref{fig:Comp} we report the stress--strain curves for the four foam families. The maximum deformation reaches $35$\% strain for the largest compression load. Beyond $35$\% strain the foams are mechanically unstable.
From Fig.~\ref{fig:Comp}, we observe that at small strain foams are in the elastic regime, and the material is characterized by a full recovery to the original shape when the load is removed. Subsequently, we find a plateau with a slope similar for all our foam families, which models the structural collapse at a nearly constant stress by bending or fracture of the building blocks.

\begin{figure}
\centering
\subfigure[A]{
\includegraphics[width=0.2\textwidth]{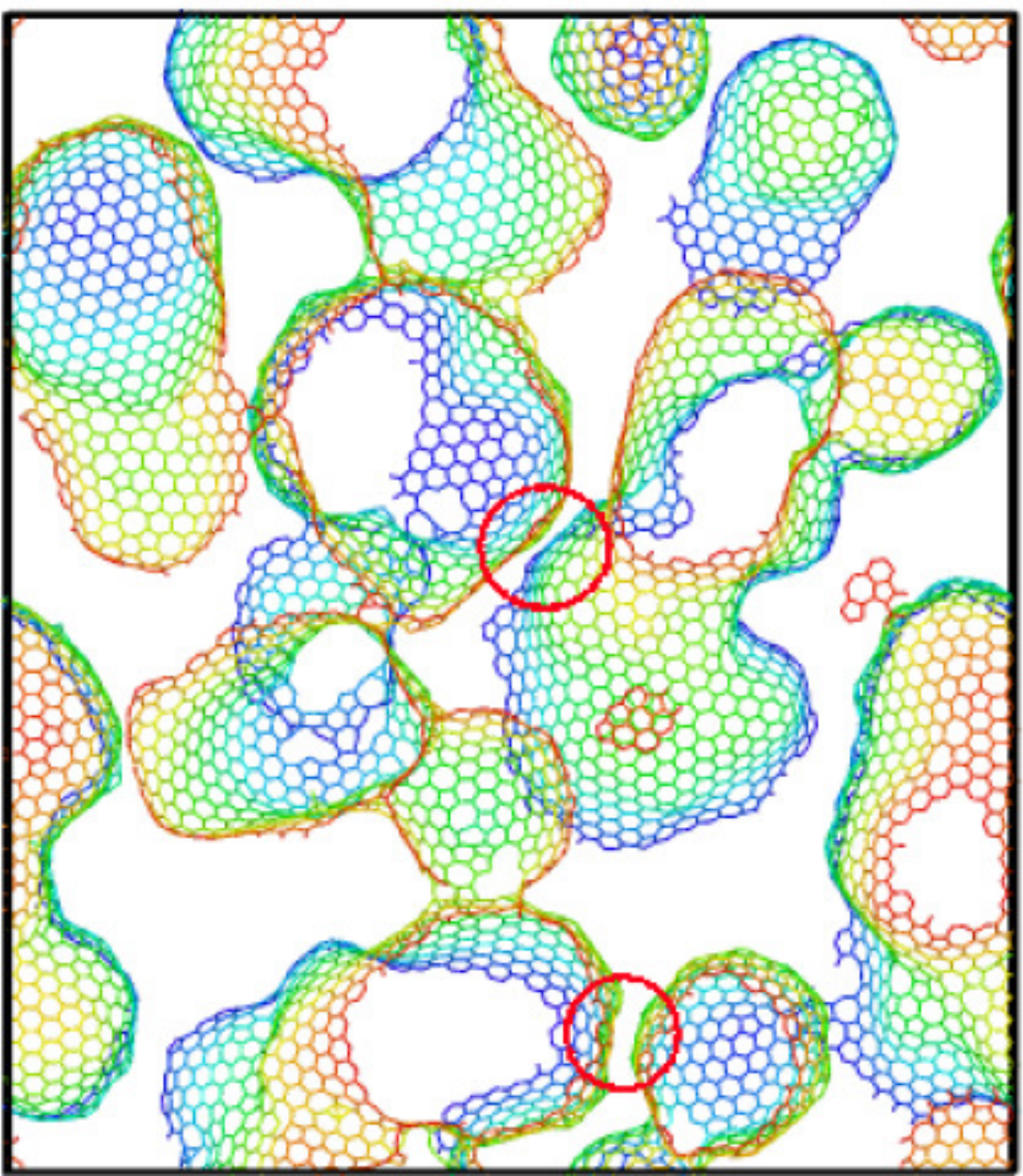}}
\subfigure[D]{
\includegraphics[width=0.2\textwidth]{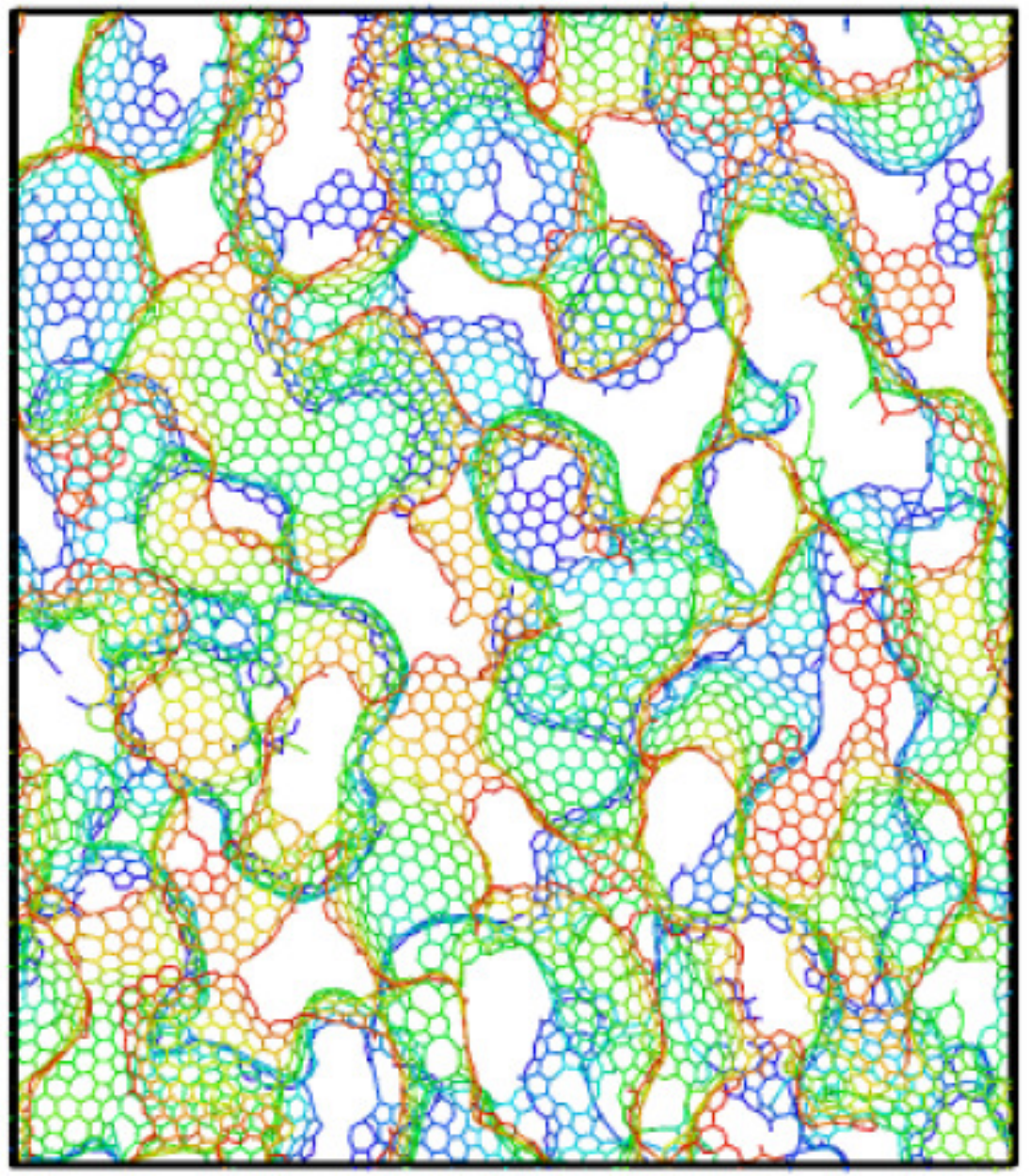}}
\caption{A $2.0$~nm slice of a sample from the foam families A and D under $12$\% compressive strain. As by Fig.~\ref{fig:Comp} this strain value sets the transition between the elastic regime and the collapsing plateau. This transition for foams of higher porosity, such as those belonging to the type A, is related to the closure of the interstitial space when graphene sheets touch upon, as those highlighted by red circles.}
\label{fig:12Comp}
\end{figure}

Finally, at higher strain (not shown) one finds a steep ramp in the stress--strain curve, representing the complete collapse of the structures. The random foam family A, characterized by the lowest density, presents this ramp at $70$\% compressive strain. At variance, higher density random foams are not stable under compression before their respective final ramps, and present a structural transition from graphene ordered layers to amorphous carbon, with a strong stress decrease followed by an increase.  

\begin{figure}[hbt]  
\centering
\includegraphics[width=0.5\textwidth]{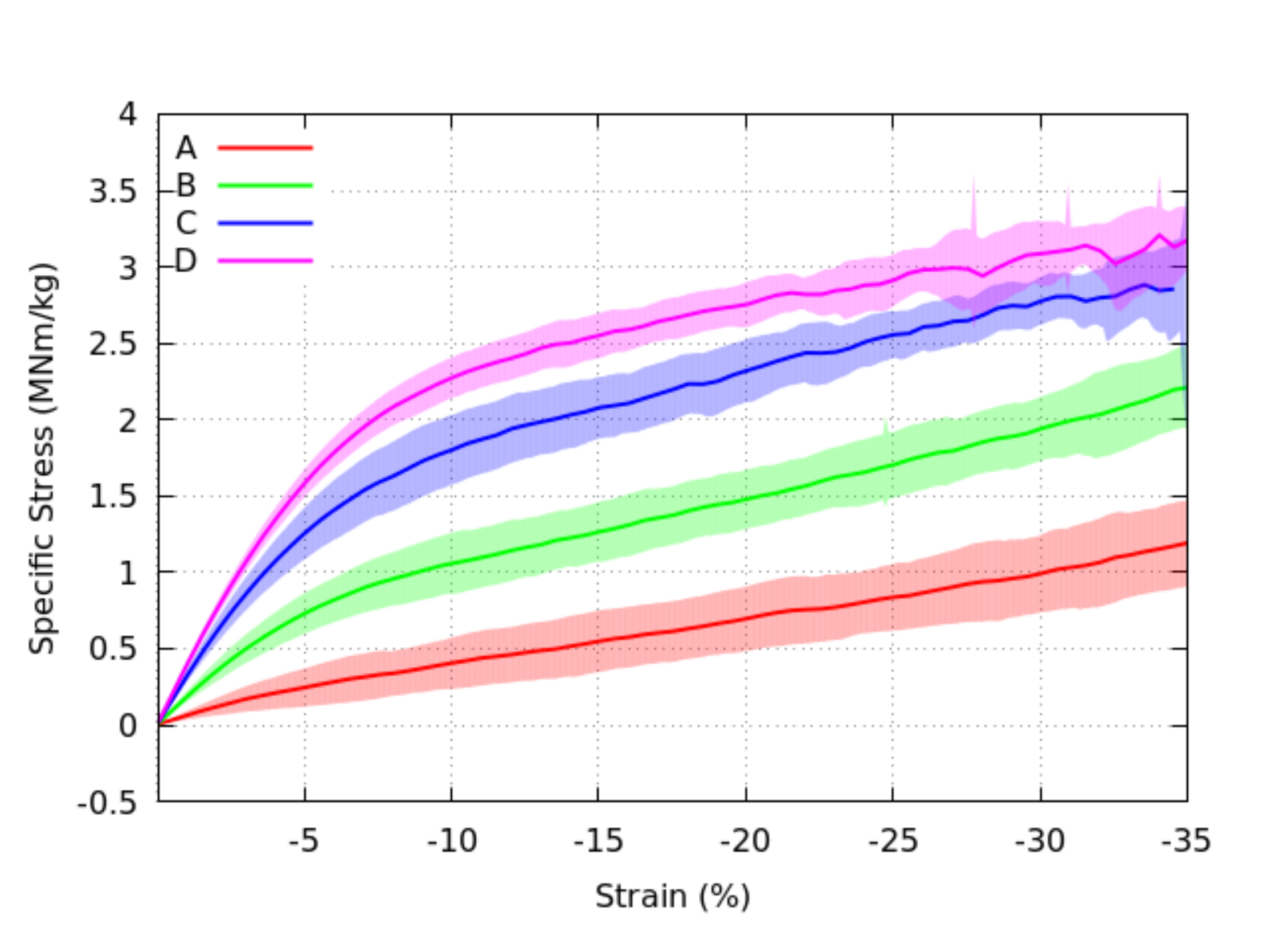}
\caption{Specific stress--strain curves of the four random foam types under uni-axial compression up to $35$\% strain. Shaded areas represent deviation within each foam family.}
\label{fig:CompSpec}
\end{figure}

The most visible mechanical characteristics of our families of random foams (see the stress--strain curves in Fig.~\ref{fig:Comp}) is that, with increasing foam density and decreasing pore size average dimension, the elastic part presents an increasing slope, while beyond $5-10$\% strain the slopes are very similar. This behavior suggests a change in the compression mechanism: below $5-10$\% strain the slope is mainly due to the connectivity among graphene layers and this regime is characterized by structural stability, while beyond that the structures start collapsing with a relatively small increase of the stress, due to the bending of the graphene sheets inside the foams.

The foam family with the lowest mass density presents an almost linear stress--strain characteristic. This suggests that the collapse is dominated by bending. In higher density foams the slope change is more marked, showing that bending of graphene sheets occurs at higher strain. 

We notice that the similarity of the slope of the stress--strain curves between $10$\% and $30$\% strain is due to a similar mechanism for collapse. This similarity can be rationalized by observing  Fig.~\ref{fig:12Comp}, where a $2.0$~nm slice of a sample from the foam families A and D under $12$\% compressive strain is reported. As by Fig.~\ref{fig:Comp} this strain value sets the transition between the elastic regime and the collapsing plateau. This transition for foams of higher porosity, such as those belonging to the type A, is related to the closure of the interstitial space when graphene sheets touch upon, as those highlighted by red circles. 

\begin{figure} 
\centering
\includegraphics[width=0.5\textwidth]{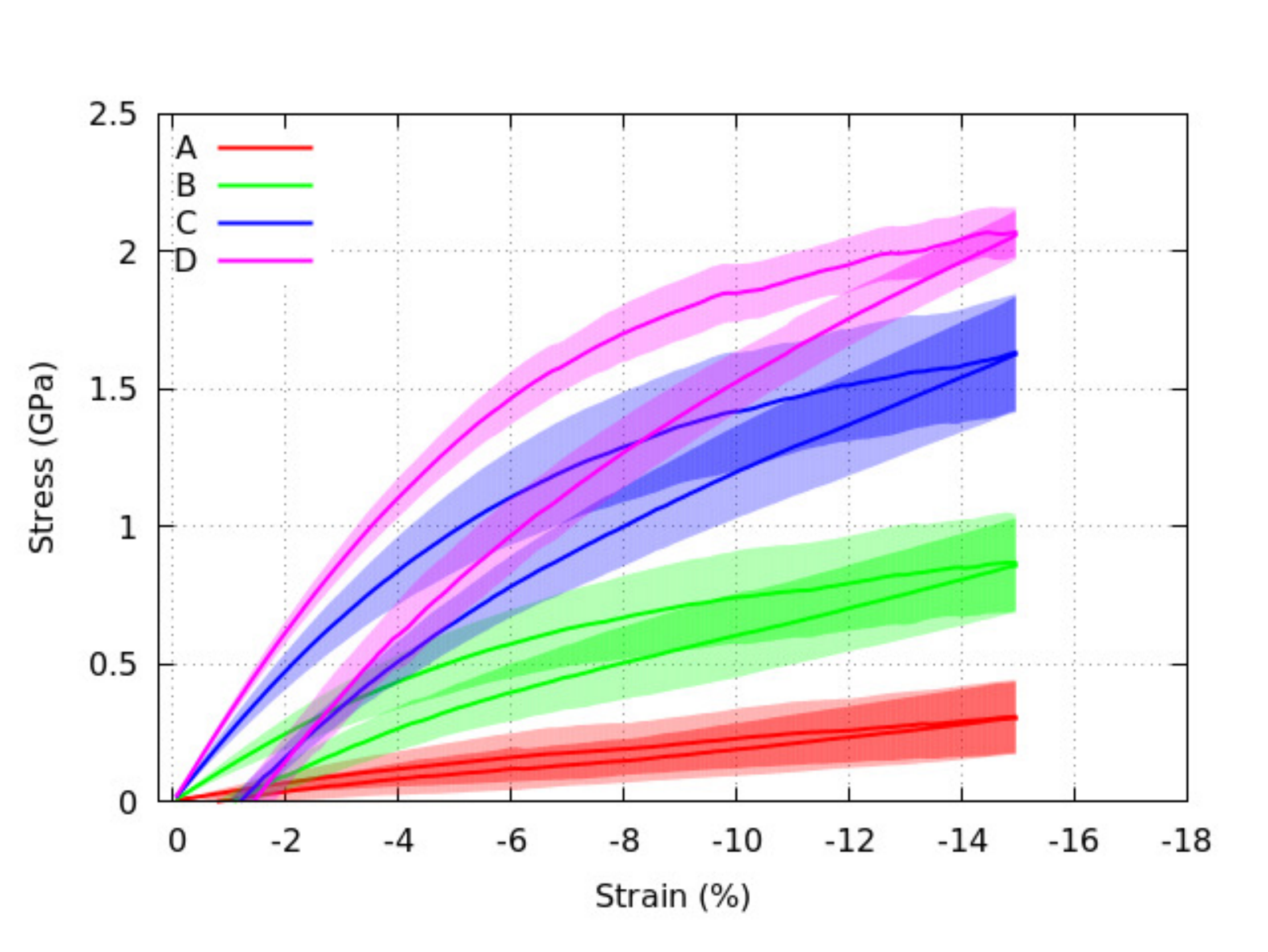}
\caption{Stress--strain curves of the four random foam types under uni-axial compression up to $15$\% strain.}
\label{fig:Comp10}
\end{figure}

In Fig. \ref{fig:CompSpec} we report (as performed in the case of tensile load) the specific stress--strain curves, which are normalized per mass density. Similarly to the tensile case, the stress--strain characteristics are almost unaffected by this normalization, once more suggesting that the mechanical performances of random foams are related to the connectivity and the topology of the foams rather than to the mass density.

Moreover, under compression the behavior of the random foam families C and D, characterized by higher density, is very similar to that of regular nanofoams presented in Ref. \citep{Pedrielli2017} with comparable plateau specific stress. At variance, lower density random foams present an almost linear behavior under compression that differs from that of regular nanofoams. Young modulus and plateau slope for the four random foam families under compressive load are reported in Tab. \ref{tab:Table C}.

\begin{table}
\centering
\small
\begin{tabular}{cccccc}
\toprule
             & Young        & Plateau &   \\ 
   Foam type   & modulus   &  slope &       \\ 
             &(GPa)        &   (GPa) &      \\
\midrule
A & 3.5  &  1.6 &    \\ 
B & 12.3 &  3.0 &   \\ 
C & 24.4 &  3.7 &   \\ 
D & 31.8 &  3.1 &  \\ 
\bottomrule
\end{tabular}
\caption{Young modulus and plateau slope of
the four foam families studied under compression.}
\label{tab:Table C}
\end{table}

Finally, in Fig.~\ref{fig:Comp10} we report the stress--strain curves of the four random foam families, initially loading the samples up to $15$\% strain and subsequently unloading them. At odds with regular foams that can fully recover their initial shape when unloaded after reaching high deformations (up to $25$\%)  \citep{Pedrielli2017}, the higher density of defects in random foams and the local concentration of stress cause an incomplete elastic behavior even for relatively small strain ($15$\%). Movies of our foams under compressive load are provided with this submission.

\subsection{Poisson Ratio}

\begin{figure}[hbtp]
\centering
\includegraphics[width=0.5\textwidth]{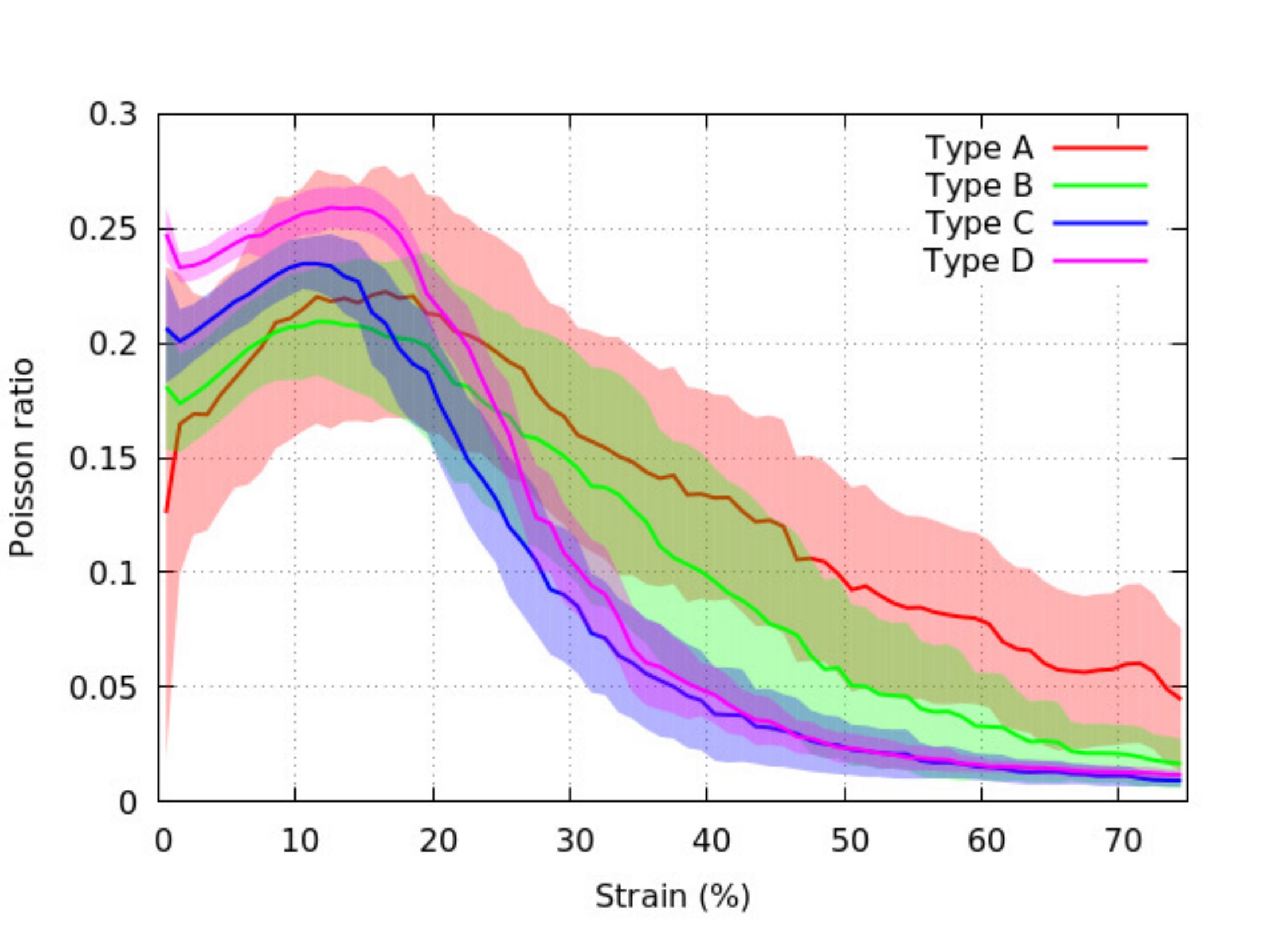}
\caption{Plot of the Poisson ratio as a function of tensile strain for the four random foam families. The shaded areas across the curves represent half of the standard deviation within each foam family.}
\label{fig:Poisson}
\end{figure}

\begin{figure}[hbt] 
\centering
\includegraphics[width=0.5\textwidth]{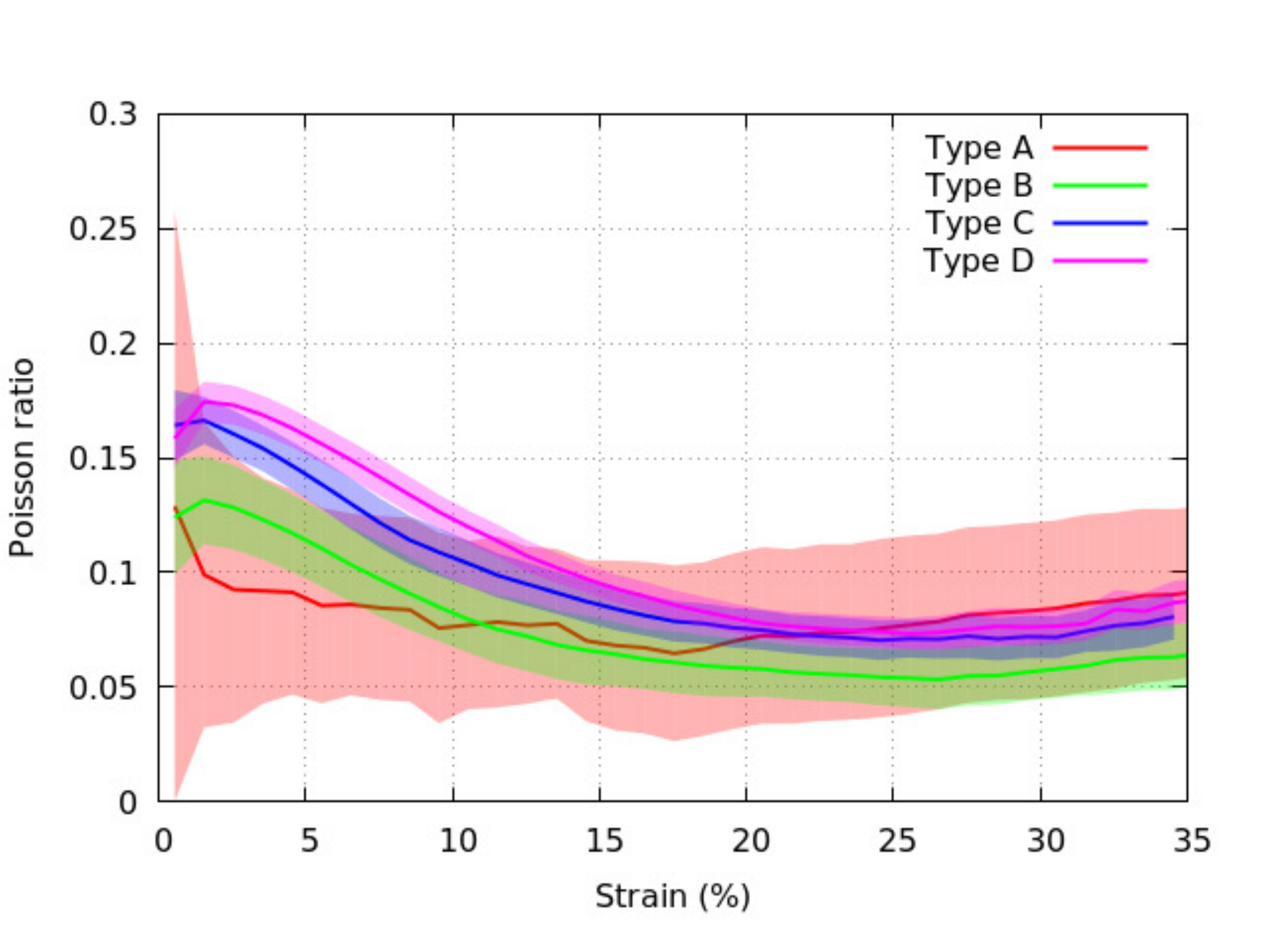}
\caption{Plot of the Poisson ratio as a function of compressive strain for the four random foam families. The shaded areas across the curves represent half of the standard deviation within each foam family.}
\label{fig:PoissonComp}
\end{figure}

To better characterize the four random foam families, we computed the Poisson ratio of these structures. The Poisson ratio for each sample is computed as the average in the two directions transverse to the loading. The Poisson ratios under uni-axial tension and compression regimes are plotted in Figs.~\ref{fig:Poisson} and \ref{fig:PoissonComp}, respectively. Notably, the Poisson ratios are positive over the whole deformation range for all our random foam families. 

In particular, for near zero strain under tension, the Poisson ratio is in the range $0.1-0.25$ with values increasing with mass density and foam connectivity. Furthermore, up to in $15$\% strain there is a small increment ($0.02-0.05$) in the Poisson ratio for all the foams families. Finally, the Poisson ratio at higher tensile strain goes to zero. This behaviour is explained as mainly due to the fracture of the samples in this deformation regime, which prevents the sample from further contraction.

At variance, under compression the decrease of the Poisson ratio is due to the internal rearrangement of the graphene layers. At higher strain (about $30$~\%) the Poisson ratio ranges between ($0.05-0.1$).

It can be worth noting that in Figs.~\ref{fig:Poisson} and \ref{fig:PoissonComp}, the standard deviation, reported as a shaded area across the relevant curve, is significantly smaller for lower density random foam families than that for higher densities.

\subsection{Thermal conductivity}

The thermal conductivity was assessed for all the samples by using the averaging procedure of the HCACF explained before. In general, HCACF dies off within $100$~ps and subsequently oscillates. This makes possible the time bead division, discussed in sec. 4.

\begin{figure}[hbt] 
\centering
\includegraphics[width=0.5\textwidth]{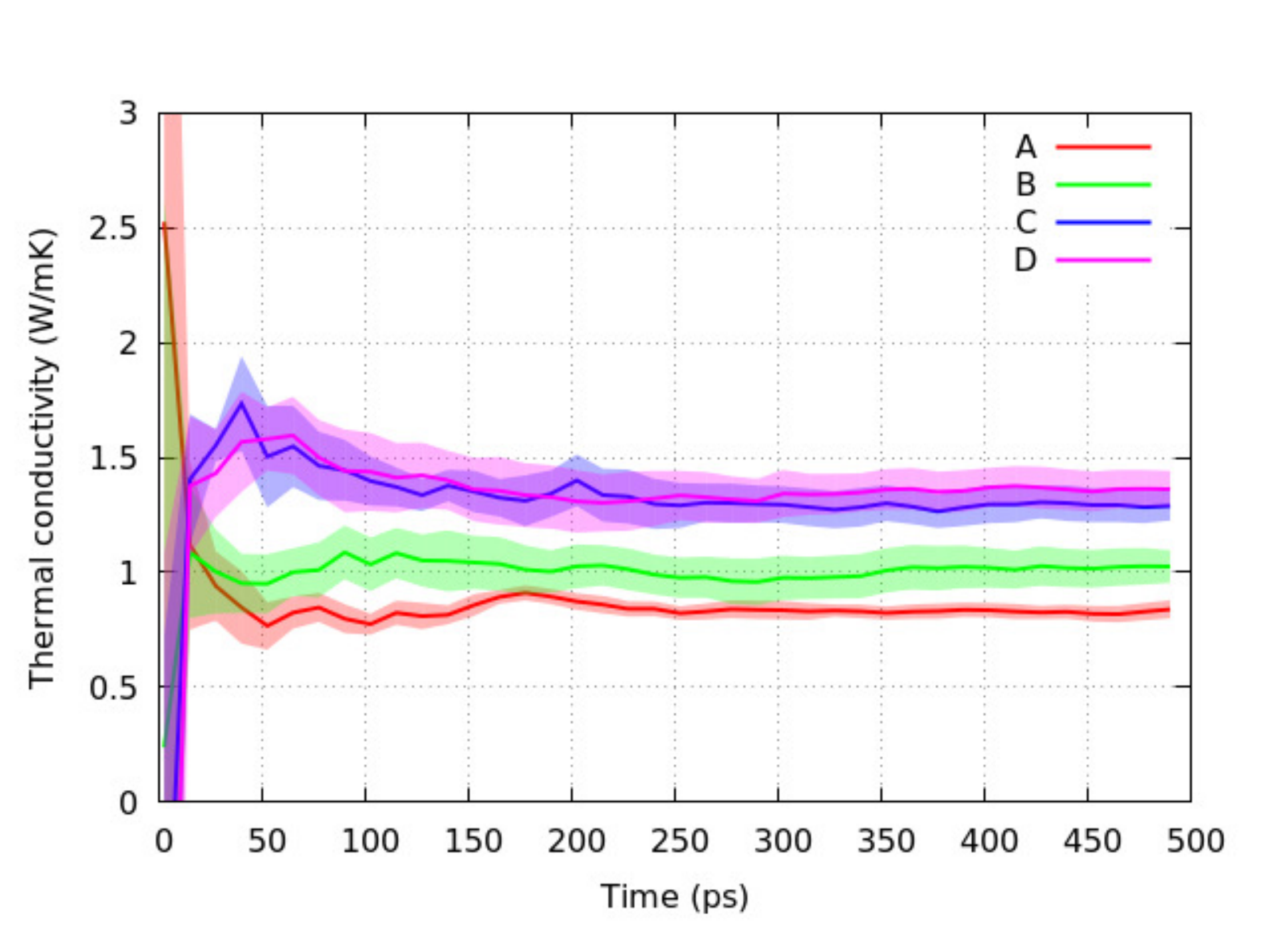}
\caption{Time averaged HCACF vs. simulation time of the four random foam families calculated as by Eq. \ref{cond}. The asymptotic values after 300 ps provide the thermal conductivity of the samples. The shaded area for each relevant curve represents half of the standard deviation.}
\label{fig:ThermalK}
\end{figure}

 In Fig.~\ref{fig:ThermalK} we plot the average of the integral of HCACF as a function of simulation time for each foam family. The thermal conductivity is given by the asymptotic values of the time-integrated HCACFs. These values for the four foam families, obtained by averaging in the range $400-500$~ps, are reported in Tab. \ref{tab:Table Thermal}. 

\begin{table}
\centering
\small
\begin{tabular}{cccccc}
\toprule
             & Thermal       & Standard  &   \\ 
   Foam type   & conductivity    &  deviation  &       \\ 
             &(Wm$^{-1}$$\cdot$ K$^{-1}$)        &    (Wm$^{-1}$$\cdot$ K$^{-1}$)    &     \\
\midrule
A & 0.83 & 0.13 &   \\ 
B & 1.02 & 0.20 &   \\ 
C & 1.29 & 0.20 &   \\ 
D & 1.36 & 0.22 &   \\ 
\bottomrule
\end{tabular}
\caption{Thermal conductivity of the four foam families computed via the Green-Kubo approach using an optimized Tersoff potential.}
\label{tab:Table Thermal}
\end{table}

\begin{figure}  
\includegraphics[width=0.45\textwidth]{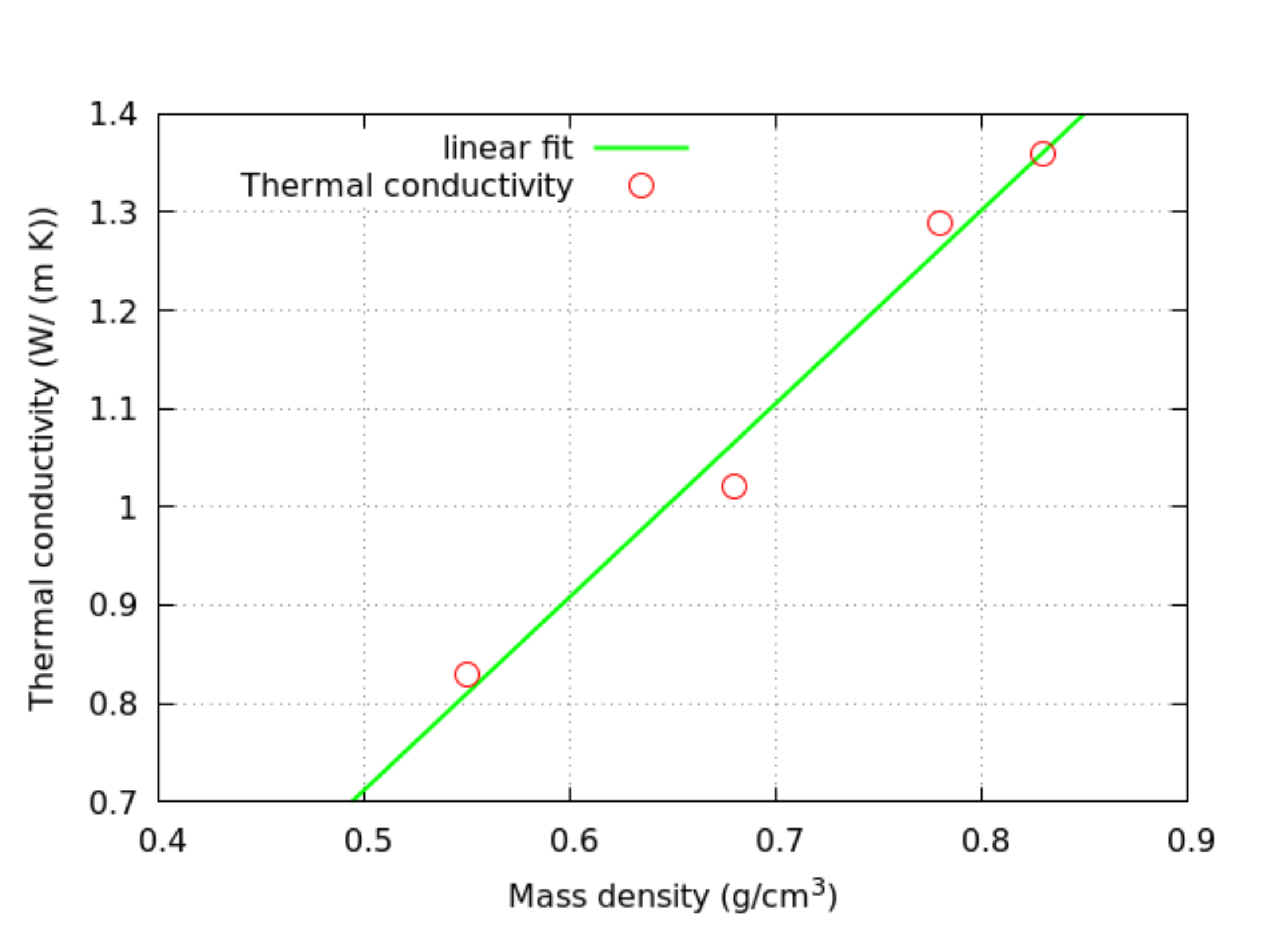}
\caption{Linear relation between mass density and the thermal conductivity for the considered foam families. }
\label{fig:ThermalLinear}
\end{figure}

We notice that two major factors affect the thermal conductivity, that are the foam connectivity and the presence of defects. In the foam families studied here, we devise that the low connectivity found in lower density foams is counterbalanced by the higher number of three-coordinated atoms, while the opposite trends occurs in higher density foams. Indeed, foams with different connectivity (see Fig. 10 of the paper) and different number of defects, show comparable values of the specific thermal conductivity (thermal conductivity per density mass unit). This could be due to a number of reasons. Our most likely explanation of this finding is as follows: smaller pore sizes generally means higher density as reported in Fig. 6. Nevertheless, a smaller size of the pores implies a larger number of defects, due to a bigger local curvature, that is a lower number of three-coordinated carbon atoms. We remind that graphene, due its particular topology, has a very large thermal conductivity, so that it is clear that a high level of carbon three-coordination is connected to large thermal conductivity. The presence of a larger number of defects in high-density foams with low-size pores should then decrease the thermal conductivity. Thus we can argue that mass density (or pore size) and number of defects counteract in the determination of the thermal conductivity. In particular, in Fig. 16 we plot the thermal conductivity as a function of the mass density for the families of four different foams, finding an almost linear relation between these two observables (coefficient of determination $R^2=0.98$).
The thermal conductivity of random foams is similar to that of glass ($1$~Wm$^{-1}$ $\cdot$ K$^{-1}$) for lower density foams, with an small increase ($1.5$~Wm$^{-1}$ $\cdot$ K$^{-1}$) for higher density foams. 

\section{Conclusions}

In this work, we investigated by means of MD simulations with reactive potentials the mechanical and thermal properties of graphene random foams with a topology experimentally achievable by growing graphene on stacked nickel nanoparticles. In particular, we tested the mechanical performances under tension and compression of four random foam families, characterized by different mass density and pore size distribution. The samples were prepared using a multi-step approach based on the Voronoi partitioning of a triangulated surface, obtained by tessellation of the simulation cell using a LJ potential forcing the carbon atoms towards a rigid support.

Under compression, we found the typical elastic deformation regime with a Young modulus significantly increasing with a decreasing average pore size dimension. A behavior, common to all the random foam families here studied, was found for compressive strain in the bending plateau zone, with a positive slope of the stress--strain curve similar for all the four foam families. For the lowest density random foam family the stress--strain characteristic is almost linear.

Finally, we calculated the Poisson ratio, a quantity used to assess the transverse response of materials to deformation, of these random foams. Under tension, the Poisson ratio is positive for all the random foam families, indicating a transverse contraction under tensile load. The values of the Poisson ratio under compression are again positive for all the considered strain and tend to stabilize as the strain increases. 

As a major outcome of our computational analysis, we find that mechanical properties under tension are characterized by an overall decrease of Young modulus with respect to regular nanofoams, while a tensile strength of the same order of that found for regular foams was obtained for higher density random foams.

Due to the interest of using foams as a mean for achieving thermal resistance, we computed the thermal conductivity of random foams using the Green-Kubo approach with a Tersoff potential optimized for these simulations. The thermal conductivity is comparable to that of glass, thus higher than materials typically used as thermal insulators, such as polyurethane rigid foams. Still, random foams do not display good thermal conductive properties, which can be related to the low connectivity in case of high porosity foams and to the presence of defects in low porosity foams. Nevertheless, combining outstanding mechanical performances with light weight, low density and good thermal insulating properties, carbon random foams could be promising candidates as reinforcing fillers in nanocomposites or elastomers to tailor their properties or to replace polymer materials in applications where thermal stability and mechanical strength are needed. 

\section*{Acknowledgements}

N.M.P. is supported by the European Commission H2020 under the Graphene  Flagship  Core  1  No.  696656  (WP14 ``Polymer composites'') and under the Fet Proactive ``Neurofibres'' No.732344.
S.T and G.G. acknowledge funding from previous WP14 ``Polymer composites'' grant.
Access to computing and storage facilities owned by parties and projects contributing to the Czech National Grid Infrastructure MetaCentrum provided under the programme ``Projects of Large Research, Development, and Innovations Infrastructures'' (CESNET LM2015042), is greatly appreciated ({\tt https://www.metacentrum.cz/en/}) The authors thanks Dr. V. Morandi and Dr. M. Christian for graphene foam SEM image and useful discussions.

\end{document}